\documentstyle[eqsecnum,epsf,aps,multicol2]{revtex}
\newcommand{\vs}{\vspace*{4mm}}

\begin{document}

\rightline{OU-TAP 92}
\centerline{\large \bf The No-Negative Mode 
Theorem}  
\centerline{\large \bf in False Vacuum Decay with Gravity } 
%\vspace*{2mm}
%\centerline{\large \bf -- No-supercritical supercurvature mode in 
%one-bubble open inflation --}
\vspace*{4mm}
\centerline{\large Takahiro Tanaka} 
\vspace*{4mm}
\centerline{\em Department of Earth and Space Science, Graduate 
School of Science}
\centerline{\em Osaka University, Toyonaka 560-0043, Japan} 

\begin{abstract}
The so-called negative mode problem in the path integral approach 
to the false vacuum decay with the effect of gravity has been an 
unsolved problem. Several years ago, we proposed a conjecture which 
is to be proved in order to give a consistent solution to the 
negative mode problem. We called it the ``no-negative mode conjecture''. 
In the present paper, we give a proof of this conjecture for rather 
general models.  Recently, we also proposed the ``no-supercritical 
supercurvature mode conjecture'' that claims the absence 
of supercritical supercurvature modes in the one-bubble open 
inflation model. In the same paper, we clarified the equivalence 
between the ``no-negative mode conjecture'' and the ``no-supercritical
supercurvature mode conjecture''. Hence, the latter is also proved
at the same time when the former is proved. 
\end{abstract}
%\pacs{} 

\vspace*{4mm}
%\centerline{\bf 27 Jan. 1999}

\multicols{2}
\section{introduction}

In recent years, 
the process of the false vacuum decay with the effect of gravity 
has been studied extensively in the context of 
the one-bubble open inflation scenario,  
in which 
an open universe is created in a nucleated bubble 
formed through false vacuum decay\cite{open}. 
In this context, 
we are interested in the spectrum of 
quantum fluctuations after the bubble nucleation because it 
determines the spectrum of primordial fluctuations 
of the universe. 
By comparing the predicted spectrum with the observed one, 
we can test whether a certain model is viable or not. 
The fluctuation can be decomposed by using the spatial harmonics 
in an open universe. 
We denote the eigen value of the spatial harmonics by $p^2$. 
The spatial harmonics with positive $p^2$ are square integrable 
on a time constant hypersurface in an open universe, and 
we have a continuous spectrum for $p^2>0$. 
While, 
the spatial harmonics are no longer square integrable when 
the eigen value $p^2$ is negative. 
However, since any time constant hypersurface in an open universe 
is not a Cauchy surface, this divergence does not directly 
prohibit the existence of such a mode. By considering 
the normalization of perturbation modes on an 
appropriate Cauchy surface, we find that 
the spectrum for $p^2<0$ can exist as a 
discrete one, which we call supercurvature mode\cite{scmode}. 

{}Further, we classified the modes with negative $p^2$ into 
two classes depending on their eigen values. 
One is supercritical modes with $p^2<-1$ 
and the other is subcritical modes with $p^2>-1$. 
If and only if there exist supercritical modes, the two-point 
correlation function of the tunneling field unboundedly increases 
for large spatial separation in the open universe created in 
the nucleated bubble.  
The diverging correlation for large spatial separation 
does not seem to be allowed intuitively. 
Furthermore, it is known that we meet a trouble 
in constructing homogeneous fluctuations 
for supercritical modes except for some restricted cases\cite{juan}. 
Hence, we proposed the ``no-supercritical supercurvature mode 
conjecture'', which is the conjecture claiming that 
the existence of supercritical supercurvature modes 
should be prohibited in realistic models of creation of 
an open universe\cite{TSnew}.

On the other hand, 
there is an unsolved problem in the 
Euclidean path integral approach to 
describe the phenomena of the true vacuum bubble nucleation 
with the effect of gravity through quantum tunneling\cite{TS92}. 
That is the so-called negative mode problem\cite{Colema}. 
In the lowest WKB approximation, the quantum tunneling 
is described by a bounce solution\cite{Colema,CalCol}.  
The decay rate per unit volume and per unit time interval,  
$\Gamma$, is given by the formula
\begin{equation}
 \Gamma={\rm Im} (K)e^{-(S_E^{(bounce)}-S_E^{(trivial)})}, 
\end{equation}
where $S_E^{(bounce)}$ is the classical Euclidean action 
for the bounce solution and $S_E^{(trivial)}$ is that 
for the trivial solution that stays in the false vacuum. 
In the path integral approach, 
the prefactor ${\rm Im}(K)$ is evaluated by taking the imaginary part 
of the gaussian integral over fluctuations around the 
background bounce solution. 
For an ordinary system without gravity, 
there is one perturbation mode in which direction the 
action does not decrease. 
It is called negative mode. 
The existence of an unique negative mode has already proved for 
quantum fields in flat spacetime\cite{Coluniq}. 
In evaluating this gaussian integral, the path of integration 
is deformed on the complex plane to make the integral 
well-defined. 
As a result, one imaginary factor, $i$, appears in 
$K$. In the Euclidean path integral approach to the 
tunneling, this imaginary unit plays 
a crucial role when we interpret $\Gamma$ as the decay rate. 

However, in the case where gravity is taken into account, 
the situation is different. 
In this case, the tunneling is described by 
the $O(4)$-symmetric Coleman and 
De Luccia (CD) bounce solution\cite{ColDeL}. 
In reducing the action for fluctuations 
around the $O(4)$-symmetric CD bounce solution, 
we used the standard gauge fixing method 
to deal with gauge degrees of freedom\cite{dirac}. 
Then, we obtained the reduced action that retains 
only the physical degrees of freedom\cite{TS92,GXMT1}. 
Reflecting the fact that the action 
with gravitational degrees of freedom is unbounded 
below, the reduced action for the fluctuations that 
conserve the $O(4)$-symmetry 
has an unusual overall signature. 
To treat this system, we proposed to use the 
prescription similar to the conformal rotation\cite{TS92,GiHaPe}. 
Then, from path integral measure, there arise one 
imaginary unit $i$. 
Hence, in order to obtain a correct number of $i$, i.e., one, 
it seems that there should exist no-negative mode.
Therefore, we proposed the ``no-negative mode conjecture''. 

We have shown in our previous paper\cite{TSnew} that  
the existence of supercritical supercurvature modes 
is equivalent to the existence of negative modes. 
That is, if we obtain a proof of the ``no-negative mode conjecture'', 
the ``no-supercritical supercurvature mode conjecture''
is also proved. 
Henceforce, the issue 
if the ``no-negative mode conjecture'' is true or not is 
now of increasing importance. 
In the same paper\cite{TSnew}, we have shown that this 
conjecture holds in a certain restricted situation. 
In the present paper, we give a proof of this conjecture 
for rather general models which consist of one scalar field. 

Here, we briefly describe the statement proved in the present paper. 
Here, 
we should note that not all CD bounce solutions contribute to the 
tunneling process. 
The relevant contribution to the decay rate 
comes only from the bounce solution 
that realizes the minimum value of action 
among all non-trivial solutions. 
Thus, the statement that we should prove is the following:
``There is no-negative mode for 
the CD bounce solution that realizes the minimum value of 
action among all non-trivial solutions.'' 
%We call this the ``weak no-negative mode conjecture''. 

However, making a list of all the 
CD bounce solutions, we find that there are many solutions 
for which the tunneling field 
does not change monotonically. 
These solutions have more than two concentric domain walls. 
We expect that the simplest bounce solution with one-domain wall  
will dominate the tunneling process. 
Hence, one may want to strengthen the conjecture as follows; 
For the bounce solution that gives the minimum value of action,  
the tunneling field $\phi$ changes monotonically, and 
this solution have no-negative mode. 
We refer to this version of conjecture as 
the ``no-negative mode theorem''.
We used the word ``theorem' instead of `conjecture' 
because this statement is proved in the this paper.

This paper is organized as follows. 
In section II, we explain the ``no-negative 
mode theorem'' in more detail. 
In section III, we present a prescription to search for all 
CD bounce solutions. 
In section IV, we give a method to count the number 
of negative modes for a given CD bounce solution. 
In section V, we give a method to count the multiplicity  
of domain walls contained in a given soltion. 
In section VI, we give a method to compare the values 
of action between various CD bounce solutions 
for a given potential model. 
In Section VII, 
combining the results obtained in Sec. IV, V and VI, 
we prove the ``no-negative mode theorem''. 
We summarize the outline of the proof and 
discuss the implication of this theorem in section VIII.

\section{statement to be proved}
\label{secNNC}

In this section, we explain the statement, which we call
the ``no-negative mode theorem'', 
in more detail to clarify what we are going to prove.

\begin{figure}[bht]
\centerline{\epsfxsize7cm\epsfbox{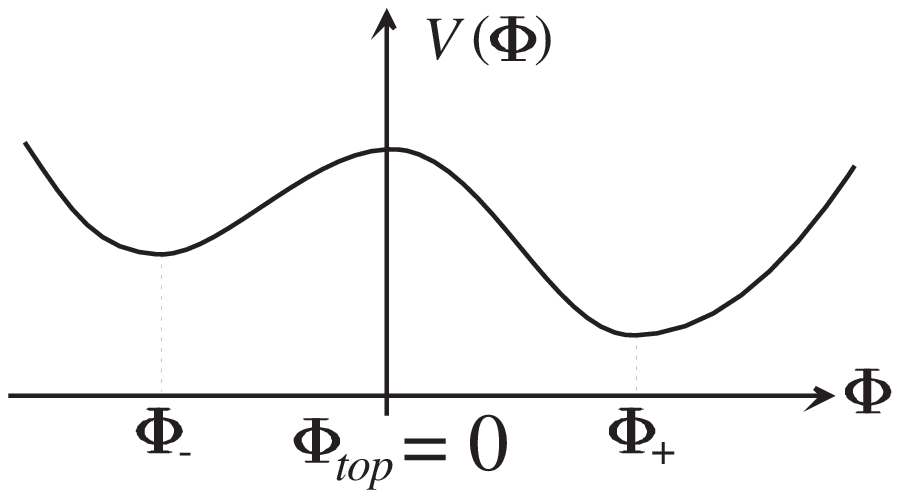}}
\vspace{3mm}
\refstepcounter{figure}
{\small FIG.\thefigure.
A typical shape of a tunneling field potential under
  consideration. We assume that the potential take its extrema 
only at $\Phi_{\pm}$ and $0$.}
\addcontentsline{lof}{figure}
{\protect\numberline{\thefigure}{fig1}}
\label{fig1}
\end{figure}

We consider the system composed of a 
scalar field, $\Phi$, with the Einstein gravity.  
The Euclidean action is given by 
\begin{equation}
S_E=\int d^4x\,\sqrt{g}\left[
-{1\over 2\kappa}R+{1\over 2} g^{\mu\nu}
\partial_{\mu}\Phi\partial_{\nu}\Phi+V(\Phi)\right],  
\end{equation}
where $\kappa=8\pi G$ and $R$ is the Ricci curvature. 
The potential of the scalar field is assumed 
to have the form as shown in Fig.\ref{fig1}, and initially 
the field is trapped in the false vacuum. 
We set $\Phi=0$ at the top of the potential barrier and 
we denote the bottoms in the false and true vacua by 
$\Phi_-$ and $\Phi_+$, respectively. 
We assume that $dV/d\Phi$ vanishes 
only at $\Phi_+$, $0$ and $\Phi_-$.

The bounce solution is an Euclidean solution 
which connects the configurations before and after tunneling. 
In the present case, before tunneling the geometry is given by 
the de Sitter space and the scalar field is in the false vacuum. 
After tunneling, there appears a true vacuum bubble in the false 
vacuum sea. 
Under the assumption of the $O(4)$-symmetry, 
\begin{eqnarray}
 &&ds^2=N^2(\tau) d\tau^2+a^2(\tau)\left\{d\chi^2
    +\sin^2\chi d\Omega_{(2)}^2\right\},
\cr
 &&\Phi=\phi(\tau), 
\end{eqnarray}
the corresponding bounce solution is found  
by Coleman and De Luccia\cite{ColDeL}. 
With this assumption, 
the Euclidean action reduces to 
\begin{equation}
 S_E=2\pi^2\int_{\tau_-}^{\tau_+} d\tau N
  a^3 \left[{1\over 2N^2}\dot\phi^2+V(\phi) -{3\over \kappa}
   \left({H^2\over N^{2}}+{1\over a^2}\right)\right], 
\label{newaction}
\end{equation}
%\vspace{5mm}
\vspace{5mm}
where $~\dot{}~$ represents the differentiation with respect to 
$\tau$ and $H:=\dot a/a$. 
Here $\tau_{\pm}$ is the value of $\tau$ 
at both boundaries. We assume 
that $\tau_+ >0$ and $\tau_- <0$. 
For definiteness, we assume that $\phi$ is 
in the true (false) vacuum side at $\tau=\tau_+\,(\tau=\tau_-)$.  
Due to this restriction, 
we do not take account of solutions in which $\phi$ is in the 
same vacuum side at both boundaries. 

The Euclidean equations of motion are obtained by 
taking the variation of $S_E$. 
After taking variations, we adopt the synchronous gauge 
by setting $N=1$, for simplicity.  
The differentiation with respect to $N$ 
gives a constraint equation, 
\begin{equation}
  H^2-{1\over a^2}={\kappa\over 3}
     \left({1\over 2}\dot\phi^2
    -V(\phi)\right). 
\label{FRWeq}
\end{equation}
From the variation with respect to $\phi$ and $a$, 
we obtain the equations of motion, 
\begin{eqnarray}
 && \ddot\phi+3H\dot\phi-{dV(\phi)\over d\phi}=0,
\label{feq}
\\
 && \dot H +{1\over a^2} = -{\kappa\over 2}\dot\phi^2, 
\label{dotH}
\end{eqnarray}
where we also used the constraint equation (\ref{FRWeq}) 
to write the equation in the form as given in (\ref{dotH}). 
We refer to the above two equations, (\ref{feq}) and (\ref{dotH}), 
as the background equations. 
Requiring the regularity at both boundaries, 
the boundary conditions to be satisfied by the background 
solutions are determined as 
\begin{equation}
 \dot\phi\to 0,\quad
 a\to \vert\tau-\tau_{\pm}\vert,
 \quad ({\rm for}\quad\tau\to \tau_{\pm}). 
\label{boundary} 
\end{equation}
Recently, Hawking and Turok proposed to relax these boundary 
conditions to allow singular instantons\cite{HawTur}. 
However, the interpretation of singular instantons is still 
an unsettled issue\cite{Alex}. 
Here we do not consider this possibility.

Now, we consider fluctuations around this bounce solution, 
to examine the prefactor $K$ that arises 
in the estimate of the decay rate. 
As briefly noted in Introduction, 
the number of $i$ in the prefactor $K$ is 
important in the path integral approach. 
To evaluate the number of $i$, we need to obtain the 
reduced action for the fluctuations 
around the bounce solution. 
Especially, $O(4)$-symmetric perturbations
\begin{eqnarray}
 ds^2 & = & \left(1+2A(\tau)\right) d\tau^2 
\cr   && +a^2\left(1+2H_L(\tau)\right) 
    \left(d\chi^2 +\sin^2\chi d\Omega_{(2)}^2 \right), 
\cr 
 \Phi & = & \phi+\delta\phi(\tau), 
\end{eqnarray}
are most important. 
After an appropriate gauge fixing, we obtain the reduced 
action for $O(4)$-symmetric perturbations as\cite{TS92,GXMT1}
\begin{equation}
 \delta^{(2)}S=-{3\over 2} 
 \int d\eta \left[i\pi^{q}{dq\over d\eta}
  +{1\over 2}(\pi^{q})^2+{1\over 2}(U-3)
  q^2\right], 
\label{action2}
\end{equation}
with the potential $U$, 
\begin{equation}
 U={\kappa\over 2}\phi'{}^2-
  {\phi'''\over \phi'}+2\left({\phi''\over \phi'}\right)^2,  
\end{equation}
where $q$ is the $O(4)$-symmetric part of 
the gauge invariant variable 
introduced in Ref.~\cite{GXMT1}, and 
$\pi^{q}$ is its conjugate momentum. 
The Euclidean conformal time $\eta$ is 
related to $\tau$ by 
$d\eta=d\tau/a(\tau)$. 
As given in Ref.~\cite{GXMT1}, perturbations in 
terms of original variables are written down in the Newton gauge 
as 
\begin{equation}
 A = {\kappa\phi'q\over 2a}, 
\quad
 H_L = -{\kappa\phi'q\over 2a}, 
\quad
 \delta\phi = {1 \over a\phi'}{d (\phi'q)\over d\eta}, 
\label{Newtong}
\end{equation}
where ${}'$ denotes the differentiation with respect to $\eta$. 

A distinguishable feature of 
the reduced action for $O(4)$-symmetric perturbations 
is that the coefficient in front of $(\pi^{q})^2$ 
becomes negative. 
Note that the prefactor $K$ is evaluated by 
\begin{equation}
 K\approx \int [d\pi^{q}\,dq]e^{-\delta^{(2)} S}. 
\end{equation}
Thus, in doing 
the integration with respect to momentum variables, 
we will find that the 
gaussian integral does not converge. 
To resolve this difficulty, we proposed to use the 
prescription similar to the conformal rotation\cite{TS92,GiHaPe}. 
By changing the variables like $\pi^{q}\to -i\pi^{q}$, $q\to iq$, 
the above integration becomes well-defined. 
If we discretize the path integral, the 
numbers of $\pi^{q}$ and $q$ integrations are different by 1. 
Therefore, this change of variable will produce one 
imaginary unit $i$. 
That is,
\begin{eqnarray}
\hspace*{-2cm} K &\approx &
 i\int [d\pi^{q}\,dq]e^{\delta^{(2)} S} 
\cr
 &\approx & i\int [dq] \exp\left[-
 {3\over 2}\int d\eta 
  \left\{{1\over 2}\left({dq\over d\eta}\right)^2
  +{1\over 2}(U-3) q^2\right\}
    \right]. 
\hspace*{-2cm}
\cr && 
\end{eqnarray}

To determine the number of $i$ in $K$, 
we also need to know the spectrum of eigen values, $\lambda_j$, 
of the following Schr\"odinger type equation: 
\begin{equation}
 \left(-{d^2\over d\eta^2}+U-3\right)q_j=\lambda_j q_j. 
\label{yeq}
\end{equation}
The contribution to $K$ from $O(4)$-symmetric perturbations 
will be given by $\approx i \displaystyle\prod^j \lambda_j^{-1/2}$. 
If there is no mode which has a negative eigen value, 
there arises no additional imaginary unit factor. 
Then, the prefactor $K$ becomes imaginary. 
Thus, we conjectured that 
there should not be any negative mode for the system 
including the gravitational degrees of freedom, 
in Ref.~\cite{TS92}. 
As discussed in Ref.~\cite{TSnew,TS92}, one can show that, 
if there is no-negative mode in 
$O(4)$-symmetric perturbations, 
it is also the case for the other perturbation modes. 
Hence, it is sufficient to concentrate on  
$O(4)$-symmetric perturbations. 

\endmulticols

%\vspace{5mm}
\begin{figure}[bht]
\centerline{\epsfxsize15cm\epsfbox{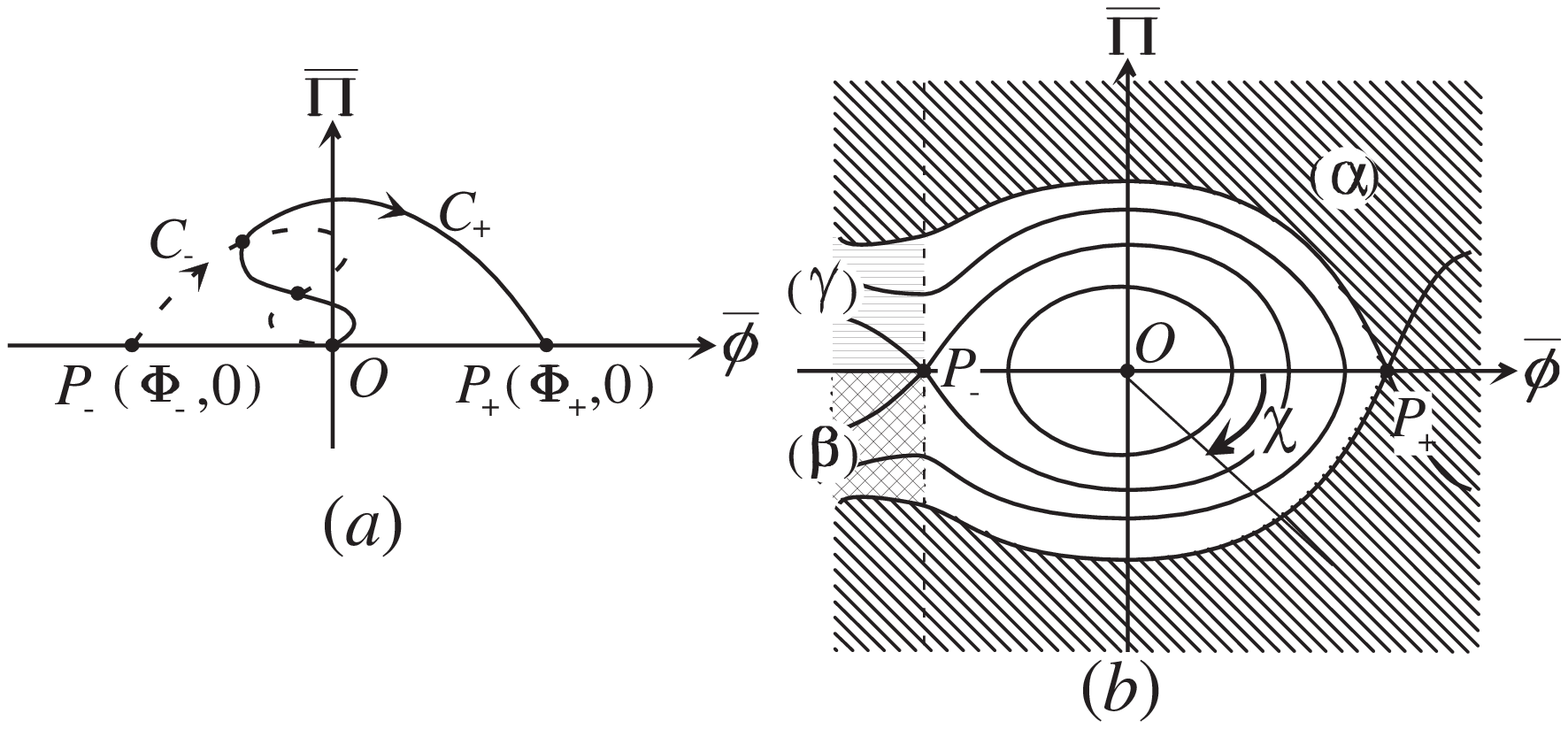}}
\vspace{3mm}
\caption{$(a)$ A schematic 
plot of $(\bar\phi_{\pm}(\phi_{\pm}^i),\bar\Pi_{\pm}(\phi_{\pm}^i))$. 
$\bar\phi_{\pm}(\phi_{\pm}^i)$ and  
$\bar\Pi_{\pm}(\phi_{\pm}^i)$ are the values of 
$\phi_{\pm}(\phi_{\pm}^i;\tau)$ and its conjugate 
evaluated at the maximum radius. 
$\phi_{\pm}(\phi_{\pm}^i;\tau)$ is obtained by solving 
the background equations with the regular boundary condition 
at $\tau=\tau_{\pm}$, where $a=0$.
The direction of the curves $C_{\pm}$ is determined so 
that $\phi_{\pm}^i$, the initial value of $\phi_{\pm}$, 
increases along the curves.
$(b)$ The thick curves are $\rho$-constant contours. 
The curves $C_{\pm}$ do not enter the shaded regions 
labeled with $(\alpha)$ and $(\beta)$.  
Also in this figure, 
the angular coordinate $\chi$ is introduced. 
Note that $\chi$ increases in the clockwise direction. 
}  
\label{fig2}
\end{figure}
%\vspace{5mm}

\multicols{2}

As we have shown in Ref.~\cite{TSnew}, 
we can construct models of CD bounce solutions 
which possess negative modes. 
Hence, we cannot prove the absence of 
negative modes for arbitrary models of CD bounce solutions. 
However, as mentioned in Introduction, 
the important point is that 
not all CD bounce solutions contribute to the 
tunneling process. 
The bounce solution that determines the decay rate 
is the one that realizes the minimum value of action 
among all bounce solutions. 
Here, we strictly define the ``no-negative mode theorem'', 
a proof of which we 
are going to present in this paper.\vs

\noindent
{\bf The Main Theorem}
(the no-negative mode theorem):\newline
{\it For 
the CD bounce solution that realizes the minimum value of 
action among all non-trivial solutions, 
$\phi'$ has a definite signature and 
the eigen value equation, Eq.$(\ref{qeq})$, 
has no negative eigen modes.}

\section{a new method to find all $O(4)$-symmetric 
CD bounce solutions}
\label{oldsec2}

We develop a method to construct a complete list 
of $O(4)$-symmetric CD bounce solutions in this section.  
For this purpose, we introduce a diagram 
consisting of two curves. We show that 
each intersection point 
of these curves corresponds to a CD bounce solution. 

In solving the background equations, 
if we fix the boundary value of $\phi$ at one side, 
say $\phi=\phi^{i}_{+}$ at $\tau=\tau_+$,  
the boundary conditions (\ref{boundary}) completely 
determines the initial condition to solve the equations of 
motion, (\ref{feq}) and (\ref{dotH}). 
Then, solving Eqs.~(\ref{feq}) and (\ref{dotH}) from both sides to 
the maximum radius, 
we obtain functions which satisfy the background 
equations in the respective half Euclidean regions. 
We denote them by 
\begin{equation}
 \phi_{\pm}(\phi^i_{\pm};\tau),\quad a_{\pm}(\phi^i_{\pm};\tau),
\end{equation}
where the subscript $+ (-)$ is attached with the variables 
that are used in solving the background equations 
from the true (false) vacuum side. 
When we solve the background equations from $\tau=\tau_+ (>0)$, 
the initial condition is varied by sweeping $\phi_+^i$ 
between the true vacuum minimum 
and the top of the potential, i.e., 
$\Phi_{top}\equiv 0\leq\phi_+^i\leq\Phi_{+}$. 
While, when we solve them from 
$\tau=\tau_- <0$, $\phi_0^i$ is sweeped between the false vacuum minimum 
and the top of the potential, i.e., 
$\Phi_-\leq\phi_-^i\leq 0$.

Here, we move the origin of the $\tau$-coordinate to   
the point corresponding to the maximum radius, 
i.e., $\dot a_{\pm}(\phi_{\pm}^i,0)=0$. 
Thus, $\tau_{\pm}$ is determined by solving 
$a_{\pm}(\phi_{\pm}^i,\tau_{\pm})=0$, and 
becomes a function of $\phi^i_{\pm}$ . 
Further, we define two functions of $\phi^i_{\pm}$ by 
the values of $\phi_{\pm}$ and $\Pi_{\pm}:=2\pi^2 a_{\pm}^3 
\dot\phi_{\pm}$ at $\tau=0$. 
We refer to 
them as $\bar\phi_{\pm}(\phi^{i}_{\pm})$ and 
$\bar\Pi_{\pm}(\phi^{i}_{\pm})$, 
respectively, where the subscript $\pm$ is used in 
the same manner as before. 

Then,  as shown in 
Fig.\ref{fig2}$(a)$, the points
$(\bar\phi_{+}(\phi^{i}_{+}),$ $\bar\Pi_{+}(\phi^{i}_{+}))$ 
and
$(\bar\phi_{-}(\phi^{i}_{-}),$ $\bar\Pi_{-}(\phi^{i}_{-}))$ 
draw two curves, $C_{+}$ and $C_{-}$, 
on the $(\bar\phi,\bar\Pi)$-plane 
as $\phi^i_+$ and $\phi^i_-$ are sweeped. 
Since $\phi_{\pm}(\phi^i;\tau)$ stays at $\phi_{\pm}^i$ when 
$\phi_{\pm}^i=\Phi_{\pm}$ or $\phi_{\pm}^i=0$, 
the curve $C_-$ starts from $P_-:=(\Phi_-,0)$ and terminates at 
$O:=(\Phi_{top}=0,0)$ and the curve $C_{+}$ starts from 
$O$ and terminates at $P_+:=(\Phi_+,0)$. 
We refer to the curves moving in the opposite direction 
as $(-C_{\pm})$.  
\vs\newline
{\bf Proposition 1}:
{\it 
Each of the curves $C_{\pm}$ does not intersect with itself}
\footnote{
Define curves $C'_{-}$ and $C'_+$ by connecting points 
$(\phi_{-}(\phi^i_{-}), -\Pi_{-}(\phi^i_{-}))$ and
$(\phi_{+}(\phi^i_{+}), -\Pi_{+}(\phi^i_{+}))$, respectively. 
By the same reasoning, we can also prove that $C_{\mp}$ does not 
intersect with $C'_{\pm}$.}.
\vs\newline
{\it Proof.}
If it were the case, we would have 
the same final values of $\phi$ and $\dot\phi$ at $\tau=0$ 
for different solutions. 
However, the background equations can be solved reversely from 
$\tau=0$, and the evolution is uniquely determined. 
This is a contradiction. \hfill$\Box$\vs
\vs\newline
{\bf Proposition 2}:
{\it 
All intersection points between $C_+$ and $C_-$  
correspond to CD bounce solutions.  
$($An exceptional case is the Hawking Moss instanton 
corresponding to $\phi^i_{\pm}=0 \cite{HawMos}.)$} 
\vs\newline
{\it Proof.}
We set $(\phi(\tau),a(\tau))=(\phi_-(\tau),a_-(\tau))$ 
for $\tau<0$ and $(\phi(\tau),a(\tau))
=(\phi_+(\tau),a_+(\tau))$ for $\tau>0$. 
Then, the values and the first derivatives of 
$\phi(\tau)$ are continuous at $\tau=0$ 
for intersection points. 
With the aid of Eq.(\ref{FRWeq}) we find 
that $a_{\pm}$ is also continuous at $\tau=0$. 
Since $(\phi_+(\tau),a_+(\tau))$ and $(\phi_-(\tau),a_-(\tau))$, 
respectively, satisfy the regular boundary conditions at $\tau=\tau_+$ 
and at $\tau=\tau_-$,  
$(\phi(\tau),a(\tau))$ satisfies those at both boundaries. 
Hence, $(\phi(\tau),a(\tau))$ 
is a CD bounce solution. \hfill$\Box$\vs

Before closing this section, we show that 
there is another general constraint on $C_{\pm}$. 
We define a function 
\begin{equation}
 \rho(\bar\phi,\bar\Pi)
  := {1\over 2}\left(-{\kappa\over 3}\rho \right)^3 
    \bar\Pi^2 -V(\bar\phi)
   ={1\over 2a^6} \bar\Pi^2 -V(\bar\phi), 
\label{rhodef}
\end{equation}
where we used Eq.(\ref{FRWeq}) with $H=0$. 
As a function of $\bar\phi$ and $\bar\Pi$, 
the definition of $\rho$ 
is implicit. To determine $\rho$ from this equation, we 
need to solve a third order algebraic equation 
like $\rho^3+\alpha^2\rho+\beta =0$, where $\alpha$ and $\beta$ 
are real. 
The left hand side of this equation is a monotonic function 
of $\rho$, and hence the real solution for $\rho$ is unique.  
We schematically plot the $\rho$-constant contours 
in Fig.\ref{fig2}$(b)$. From Eq.~(\ref{feq}), 
we can see that $E_{\pm}(\tau):={(1/ 2)}\dot\phi_{\pm}^2 -V(\phi_{\pm})$ 
decreases monotonically 
in the direction for $a_{\pm}$ to increase. 
Thus, $\rho_{\pm}:=\rho(\bar\phi_{\pm},\bar\Pi_{\pm})=E_{\pm}(0)$ 
is bounded by 
\begin{equation}
 \rho_{\pm}<E_{\pm}(\tau_{\pm})=-V(\phi_{\pm}^i)
 <-V(\Phi_{\pm}),\quad (\hbox{for}\quad C_{\pm}). 
\label{rhobound}
\end{equation}
%\vs\newline 
{\bf Proposition 3}:
{\it
The curves $C_{\pm}$ do not enter 
the shaded regions labeled with $(\alpha)$ and $(\beta)$ 
in Fig.\ref{fig2}$(b)$.}
\vs\newline
{\it Proof.}
It is manifest that $C_-$ cannot go into the regions 
$(\alpha)$ and $(\beta)$ in Fig.\ref{fig2}$(b)$ 
from the bound (\ref{rhobound}). 
On the other hand, from (\ref{rhobound}) alone, 
it seems possible that the curve $C_+$ enters 
into the region $(\beta)$.
% though the region $(\alpha)$ 
%is forbidden to be entered. 
However, for the potential that we assumed, 
$dV(\phi)/d\phi$ stays negative
once $\phi_+$ becomes smaller than $\Phi_-$. 
Hence, when we solve equations of motion 
from $\tau=\tau_+$, $\phi_+$ never stops on the left side 
of $\Phi_-$. 
Therefore, when $\bar\phi_+<\Phi_-$, 
$\bar\Pi_+$ must be positive. \hfill$\Box$\vs

For the later convenience, we introduce a new coordinate, 
$(\rho, \chi)$, where $\rho$ is given by (\ref{rhodef}) and 
$\chi$ is defined by 
\begin{equation}
 (\cos\chi,\sin\chi)={(\bar\phi,-\bar\Pi)\over\sqrt{\bar\phi^2+\bar\Pi^2}}. 
\end{equation}
for the unshaded region in Fig.\ref{fig2}$(b)$. 
For the region $(\gamma)$, which the curve $C_+$ can enter, 
we extend the $\chi$-coordinate so that it change 
monotonically along each $\rho$-constant curve and 
does not have a point with $\chi=m \pi$ in this region, 
where $m$ is an integer.  

\section{number of negative modes}
\label{oldsec5}

In this section, we develop a method to count 
the number of negative modes of Eq.~(\ref{yeq}) 
from the topology of the $(\bar\phi,\bar\Pi)$-diagram 
introduced in the preceding section. 
Unless $\phi'$ goes to $0$, Eq.~(\ref{yeq}) is a 
Schr\"odinger type equation with a regular potential. 
In this case, the number of nodes of $q_j$ for 
$\lambda_j=0$ gives the number of negative modes. 
Hence, as long as such nodeless bounce solutions 
are concerned, we can say that there is no-negative mode 
when $q_j$ for $\lambda_j=0$ has no node. 
Therefore, we concentrate on the case with $\lambda_j=0$. 
Although the above statement 
does not hold any longer once a node of 
$\phi'$ appears, we do not assume $\phi'\ne 0$ 
in most of the following discussions 
in this section. The only exception is 
the subsection C. 

In subsection A, we introduce   
variables $q_{\pm}(\tau)$ defined by Eq.(\ref{qeq}) 
with Eq.(\ref{varphi}). They are constructed from the 
background solution $\phi_{\pm}(\phi_{\pm}^i,\tau)$, 
and we show that they satisfy Eq.~(\ref{yeq}) 
for $\lambda_j=0$ with $\phi=\phi_{\pm}$. 
Then, in subsection B, we give a method to 
count the number of nodes of the functions 
$x_{\pm}:=(\dot\phi_{\pm} q_{\pm})$ from the 
$(\bar\phi,\bar\Pi)$-diagram. 
Further, we define a topological number $N$ 
assigned for each intersection point, and show 
that it should be non-negative. 
In subsection C, we prove the following 
Theorem.\vs\newline 
{\bf Theorem 1}:\newline
{\it For an $O(4)$-symmetric CD bounce solution which does not 
have a point at which $\phi'=0$, the number of negative modes 
is given by $N$ defined by Eq.$(\ref{numN})$. 
$($See also the remark given at the end of this section.$)$}

\subsection{construction of a zero mode solution}
\label{subN1}

We begin this subsection by pointing out that  
Eq.~(\ref{yeq}) for $\lambda_j=0$ is nothing but 
the perturbed equations of motion 
that are obtained by taking the variation of 
the reduced action. 
Since the $O(4)$-symmetry is the symmetry 
that the background solution possesses, 
it is natural to consider that the case with $\lambda_j=0$ 
is related to the background solutions. 
\vs\newline
{\bf Proposition 4}:
{\it We define functions $q_+(\tau)$ and $q_-(\tau)$ by 
\begin{equation}
 q_{\pm}={a_{\pm}^2\over 3\dot\phi_{\pm}}
  \left(\ddot\phi_{\pm}\varphi_{\pm}
  - \dot\phi_{\pm}\dot\varphi_{\pm}\right), 
\label{qeq}
\end{equation}
where
\begin{equation}
 \varphi_{\pm}(\tau):={\partial\phi_{\pm}(\phi_{\pm}^i;\tau)\over
\partial\phi_{\pm}^i}. 
\label{varphi}
\end{equation} 
Then, $q_{\pm}(\tau)$ satisfies Eq.$(\ref{yeq})$ 
for $\lambda_j=0$ with $\phi(\tau)=\phi_{\pm}(\tau)$
regular boundary conditions.}
\vs\newline
{\it Proof.}
First, we derive the equation 
satisfied by $\varphi_{\pm}$. 
Hereafter, we sometimes suppress the subscript $\pm$ 
to keep the simplicity of notion. 
It is obtained by 
differentiating the background equations with respect to 
$\phi^i$. From Eq.(\ref{feq}), we have 
\begin{equation}
 \ddot\varphi+3{\partial H\over \partial\phi^i}
 \dot\phi+3H\dot\varphi-{d^2 V\over d\phi^2}\varphi=0. 
\label{dphi1}
\end{equation}
From Eqs.(\ref{FRWeq}) and (\ref{dotH}), we have 
\begin{equation}
 \dot H+H^2=-{\kappa\over 3}\left(\dot\phi^2+V\right).  
\end{equation}
Differentiating this equation with respect to $\phi^i$, we obtain 
\begin{equation}
 {d\over d\tau}\left({\partial H\over \partial\phi^i}\right) 
 +2H{\partial H\over \partial\phi^i}=-{\kappa\over 3}\left(2\dot\phi\dot\varphi
    +{dV\over d\phi}\varphi\right). 
\label{dphi2}
\end{equation}
By eliminating $\partial H/\partial\phi^i$ from 
Eqs.~(\ref{dphi1}) and (\ref{dphi2}), 
we obtain a third-order differential equation 
with respect to $\varphi$. 
After a lengthy but straightforward calculation,  
one can verify that the equation that $\varphi$ satisfies 
reduces to 
\begin{equation}
 \left(-{d^2\over d\eta^2}+U-3\right)
   {a^2\over 3\dot\phi}
  \left(\ddot\phi\varphi- \dot\phi\dot\varphi\right)=0, 
\label{eqvarphi}
\end{equation}
where we used the relation 
\begin{equation}
  {\ddot H\over 2}=\left(\dot H+{1\over a^2}\right) \left(
       {\ddot\phi\over \dot\phi}\right)+{H\over a^2}. 
\end{equation}
Form Eq.(\ref{eqvarphi}), it is manifest that 
$q_{\pm}$ defined by (\ref{qeq}) satisfies 
Eq.~(\ref{yeq}) for $\lambda_j=0$ with $\phi=\phi_{\pm}$. 
We note that $q_{\pm}$ is defined not only for 
intersection points corresponding to 
CD bounce solutions but also for 
any points on the curves $C_{\pm}$. 

Next, we show that $q_{\pm}$ satisfies the required 
boundary condition, i.e., $q_{\pm}\propto (\tau-\tau_{\pm})$ 
for $\tau\to\tau_{\pm}$. 
From the boundary condition for the background solutions, 
we have
\begin{eqnarray}
 &&\phi_{\pm}\approx \phi_{\pm}^i\mp {1\over 2}
    c_{1\pm}(\tau-\tau_{\pm}(\phi^i))^2, 
\cr 
 && a_{\pm}\approx \mp c_{2\pm}(\tau-\tau_{\pm}(\phi^i)),   
\end{eqnarray}
in the $\tau\to\tau_{\pm}$ limit, 
where $c_{1\pm}$ and $c_{2\pm}$ are positive constants 
depending on $\phi_{\pm}^i$. 
By substituting these expressions into Eq.~(\ref{qeq}), 
we can show that 
\begin{equation}
 q_{\pm}(\tau)\to {c_{2\pm}^2\over 3}(\tau-\tau_{\pm}) , \quad 
   (\hbox{for}\quad \tau\to\tau_{\pm}). 
\end{equation}
Thus, $q_+(\tau)$ and $q_-(\tau)$ 
satisfy the appropriate boundary condition. \hfill$\Box$\vs

As a bonus, we also find that 
\begin{equation}
  \mp q_{\pm}(\tau) >0, \quad 
   (\hbox{for}\quad \tau\to\tau_{\pm}). 
\label{mpq}
\end{equation}

Here, we show that the expression (\ref{qeq}) 
can also be derived by considering the 
relation between the gauge invariant variable $q$ 
and $\delta\phi$ in the synchronous gauge, 
which we refer to as $\delta\phi_s(\tau)$. 
Here we note that we used the synchronous gauge 
in solving the background equations. 
Considering the case in which $q$ satisfies Eq.(\ref{yeq}) 
for $\lambda_j=0$, i.e., $q''=(U-3)q$, 
we give a gauge invariant representation of $q$ 
in terms of the original variables. 
Since the gauge transformation $\tau\to\tau+\delta\tau$ 
acts on $O(4)$-symmetric perturbations as 
\begin{equation}
 \delta_g A=-\dot {\delta\tau},\quad 
 \delta_g (\delta\phi)=-\dot\phi\, \delta\tau, 
\end{equation}
we find that 
\begin{equation}
 {a^2\dot\phi\over 3}
  \left(A-{d\over d\tau}\left({\delta\phi\over\dot\phi}\right)\right), 
\end{equation}
is a gauge invariant combination. 
By using the expressions in the 
Newton gauge (\ref{Newtong}), we can verify that this 
is the gauge invariant expression of $q$ in terms of the 
original variables. 
Henceforce, we finally find that $q$ is related to $\delta\phi_s$ as
\begin{equation}
 q={a^2\over 3\dot\phi}
  \left(\ddot\phi\,\delta\phi_s- 
   \dot\phi\,\dot{\delta\phi_s}\right). 
\label{phi2q}
\end{equation}

\subsection{number of nodes of $x_{\pm}$}
\label{subN2}

We develop a method 
to count the number of nodes of $x_{\pm}:=(\dot\phi_{\pm}q_{\pm})$, 
where $q_{\pm}$ is defined by Eq.~(\ref{qeq}). 
For the convenience, we use $x_{\pm}$ instead 
of $q_{\pm}$, but 
the number of nodes of 
$x_{\pm}$ equals to that of $q_{\pm}$ 
if $\dot\phi_{\pm}$ stays positive. \vs\newline 
{\bf Proposition 5}:
{\it 
As $\phi_{\pm}^i$ is varied along $(\mp C_{\pm})$, 
the number of nodes of $x_{\pm}(\tau)$ 
$($existing between $0$ and $\tau_{\pm}$ 
excluding the point $\tau=\tau_{\pm})$ changes 
only when $\bar x_{\pm}(\phi_{\pm}^i):=\dot\phi_{\pm}(0) 
q_{\pm}(\phi_{\pm}^i;0)=0$. 
 }
\vs\newline{\it Proof.} 
It is sufficient to prove that the location of 
a zero point of  $x_{\pm}(\tau)$ 
should change continuously 
when $\phi_{\pm}^i$ is sweeped.
To show this, we suppose the opposite case. 
If a zero point of 
$x_{\pm}(\tau)$ suddenly appeared 
at a point $\tau=\tau_c$ other than 
$\tau=0$, $\displaystyle {dx_{\pm}/d\tau}$ 
would have to vanish there 
due to the continuity of $x_{\pm}(\tau)$ with 
respect to $\phi_{\pm}^i$. 
Except for the case that $\dot\phi_{\pm}$ vanishes at $\tau=\tau_c$, 
the conditions that $q_{\pm}(\tau_c)=\dot q_{\pm}(\tau_c)=0$ 
imply $q_{\pm}\equiv 0$, and hence $x_{\pm}\equiv 0$ 
because $q_{\pm}$ satisfies a regular second order 
differential equation. 
Hence, the remaining possibility is that  
$\dot\phi_{\pm}$ vanishes at $\tau=\tau_c$. 
Then, let us examine the power series 
solution of $q_{\pm}$ around the 
singular point $\tau_c$. Near the singularity, 
the behavior of $\phi'_{\pm}$ is given by $c(\eta-\eta_c)$, 
where $\eta_c$ is the value of $\eta$ corresponding to 
$\tau=\tau_c$, and $c$ is a constant. 
Then, the potential $U$ with $\phi=\phi_{\pm}$ in Eq.(\ref{yeq}) 
behaves as $U\approx 2/(\eta-\eta_c)^2$. Hence, 
the point $\eta=\eta_c$ is a regular singular point, 
and the exponents are $-1$ and $2$. 
Thus, we find that 
the general power series solution for $x_{\pm}$ around this 
singular point is given by 
$\dot\phi_{\pm} q_{\pm}=c_1(1+\cdots)+ c_2((\eta-\eta_c)^3+\cdots)$, 
where $c_1$ and $c_2$ are constants.  
Since the case $x_{\pm}(\tau_c)=0$ is considered, 
$c_1$ must be zero. 
Therefore, $x_{\pm}$ must change its signature 
at $\tau=\tau_c$. Hence $\tau=\tau_c$ cannot be the first 
point at which the zero point suddenly appeared. 
\hfill$\Box$\vs

To represent the condition $\bar x_{\pm}=0$ in terms of 
the directions of $C_{\pm}$, 
\begin{equation}
(\bar\varphi_{\pm},\bar\pi_{\pm}^{\varphi})
:=\left({d\bar\phi_{\pm}\over d\phi_{\pm}^i},
{d\bar\Pi_{\pm}\over d\phi_{\pm}^i}\right),
\end{equation} 
we first calculate $\bar\pi^{\varphi}$ as 
\begin{equation}
 \bar\pi^{\varphi}= \bar a^3\left(
  1+\bar a^2{\kappa\dot{\bar\phi}^2\over 2}\right)\dot{\bar\varphi} 
   -{\kappa \over 2}\bar a^5 {dV(\bar\phi)\over d\bar\phi}
       \dot{\bar\phi}\bar\varphi, 
\label{pim}
\end{equation}
where we used Eq.(\ref{FRWeq}) with $H=0$. 
From this equation, we can express $\dot{\bar\varphi}$ as a function 
of $\bar\varphi$ and $\bar\pi^{\varphi}$. 
Substituting this into Eq.~(\ref{qeq}), we obtain
\begin{equation}
 \bar x={1\over 3\bar a}
   \left(1+\bar a^2{\kappa\dot{\bar\phi}^2\over 2}\right)^{-1}
   \left(\bar a^3 {dV(\bar\phi)\over d\bar\phi}
       \bar\varphi-\bar a^{-3}\bar\Pi
        \bar\pi^{\varphi}\right). 
\label{barq}
\end{equation}
By using Eq.~(\ref{rhodef}), 
we find the above equation can be written as 
\begin{equation}
 \bar x=-{\bar a^2\over 3}
 \left({\partial\rho\over \partial\bar\phi}\bar\varphi
 +{\partial\rho\over \partial\bar\Pi}\bar\pi^\varphi\right)
  =-{\bar a^2\over 3}
      {d\rho\over d\phi^i}. 
\end{equation}

{}For the convenience, 
using the $(\rho,\chi)$-coordinate introduced at the end of 
Sec.\ref{oldsec2}, 
we define angles $\Theta_+$ and $\Theta_-$ 
 by 
\begin{eqnarray}
 (\cos\Theta_{\pm},\sin\Theta_{\pm})
 =\displaystyle
 {\left({d\chi_{\pm}(\phi_{\pm}^i)\over d\phi_{\pm}^i},
   \pm{d\rho_{\pm}(\phi_{\pm}^i)\over d\phi_{\pm}^i}\right)\over
 \sqrt{\left({d\chi_{\pm}(\phi_{\pm}^i)\over d\phi_{\pm}^i}\right)^2
   +\left({d\rho_{\pm}(\phi_{\pm}^i)\over d\phi_{\pm}^i}\right)^2}}, 
\label{Thetadef}
\end{eqnarray}
along $-C_{+}$ and $C_{-}$, respectively. 
Here, $(\rho_{+}(\phi_{+}^i),\chi_{+}(\phi_{+}^i))$ 
and $(\rho_{-}(\phi_{-}^i),\chi_{-}(\phi_{-}^i))$ 
represent points corresponding to 
$(\bar\phi_+(\phi_+^i),\bar\Pi_+(\phi_+^i))$ 
and $(\bar\phi_-(\phi_-^i)$ $,\bar\Pi_-(\phi_-^i))$ 
in the $(\rho,\chi)$-coordinate, respectively. 
To remove the ambiguity in $\chi_{\pm}$ modulo $2\pi$, 
we set $\chi_{+}(\Phi_+)=0$ and $\chi_{-}(\Phi_-)=-\pi$.  

When the curves $C_-$ and $(-C_+)$ start from the vacua, 
both $q_{+}(\tau)$ and $q_{-}(\tau)$ have no node, 
the proof of which is presented in the appendix.  
Furthermore, in the $\tau\to\tau_{\pm}$ limit, 
it is easy to see $\dot\phi(\tau)> 0$, and 
$\mp q_{\pm}(\tau)>0$ was shown in Eq.~(\ref{mpq}). 
Combining these facts, we find 
that both $-\bar x_{+}$ and $\bar x_{-}$ are also 
positive when the curves $C_-$ and $(-C_+)$ 
first start from the vacua. 
Then $\Theta_+$ and $\Theta_-$ are defined without 
any ambiguity by choosing them to be continuous 
along the curves $(\mp C_\pm)$ 
with the condition that 
\begin{equation}
 0<\Theta_{\pm}<\pi,\quad (\hbox{for}\quad 
      \phi^i\to \Phi_{\pm}).
\end{equation} 
In terms of $\Theta_{\pm}$, 
the point at which the number of nodes of $q_{\pm}$ 
changes is specified by 
$\Theta_{\pm}=\pi n$, where $n$ is an integer. 

To determine whether the number of nodes increases or decreases 
at a point with $\Theta_{\pm}=\pi n$, 
we examine the signature of $\dot{\bar x}$. 
It is calculated from Eq.~(\ref{qeq}) 
with the aid of Eqs.(\ref{feq}) and (\ref{dphi1}) as 
\begin{equation}
 \dot{\bar x}
    =\left(1+\bar a^2{\kappa\dot{\bar\phi}^2\over 2}\right)
    {\dot{\bar\phi}}\bar\varphi, 
\label{xbardot}
\end{equation}
where we used $\bar x=0$. 
Then, from Fig.\ref{fig2}$(b)$, we find 
$\dot{\bar x}$ is positive 
when the curve $C_{-}$ or $(-C_+)$ 
touches a $\rho$-constant curve with its direction 
pointing in the $\chi$-direction.  
By comparing this with Eq.(\ref{Thetadef}), 
the defining eqaution of $\Theta_{\pm}$, 
we find that $\dot{\bar x}$ is positive 
when $\Theta_{\pm}=2 n\pi$, while 
$\dot{\bar x}$ is negative 
when $\Theta_{\pm}=(2 n+1)\pi$, where $n$ is an integer. 
To summarize, we have 
\begin{eqnarray}
 && \mbox{sgn}\bar x_{\pm}=
 -\mbox{sgn}\left({d\rho_{\pm}(\phi_{\pm}^i)\
              \over d\phi_{\pm}^i}\right)=
 \mp\mbox{sgn}(\sin \Theta_{\pm})\cr
  && \mbox{sgn}\dot{\bar x}_{\pm}=
 \mbox{sgn}\left({d\chi_{\pm}(\phi_{\pm}^i)\
              \over d\phi_{\pm}^i}\right)=
 \mbox{sgn}(\cos \Theta_{\pm}). 
\end{eqnarray}
%\vs\newline
%\vspace{5mm}
\begin{figure}[htb]
\centerline{\epsfxsize6cm\epsfbox{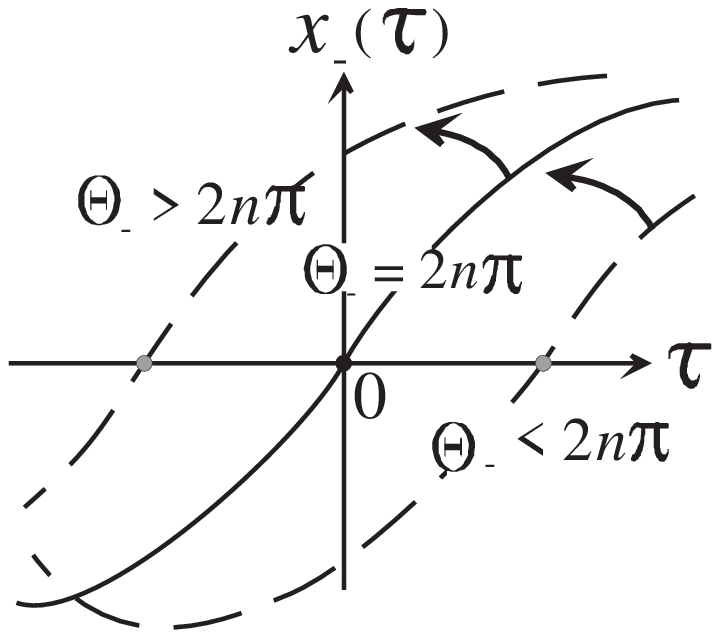}}
\vspace{3mm}
\refstepcounter{figure}
{\small FIG.\thefigure.
The behavior of $x_{-}(\phi^i;\tau)$ when 
$\Theta_-(\phi^i)\approx 2n\pi$. In this case, since 
$\dot {\bar x}_->0$, the zero point of $x_-(\tau)$ moves 
left as $\Theta_-$ increases.}
\addcontentsline{lof}{figure}
{\protect\numberline{\thefigure}{fig3}}
\label{fig3}
\end{figure}
\vspace{5mm}
{\bf Proposition 6}:
{\it we refer to the number of nodes of $x_{\pm}(\tau)$ as $N_{\pm}$. 
Then $N_{\pm}$ is determined by } 
\begin{equation}
\pi N_{\pm}<\Theta_{\pm}\leq \pi(N_{\pm}+1).  
\end{equation}
{\it Here we do not count a zero point at 
$\tau=0$ as a node even if it exists.} 
\vs\newline
{\it Proof.}
Let us consider the $\tau<0$ side. 
When $\Theta_{-}\approx 2 n\pi$, 
$\bar x_-\approx \sin\Theta_-$ changes its 
signature from $-$ to $+$ for increasing $\Theta_-$. 
Remember that $\dot{\bar x}_->0$ in this case. 
Then, it will be easy to see 
from Fig.\ref{fig3} that 
the number of nodes, $N_-$, increases by one when $\Theta_-$ 
crosses $2n\pi$ in its increasing direction. 
When $\Theta_{-}\approx(2 n+1)\pi$, $\dot{\bar x}_-$ is negative. 
In this case, $\bar x$ changes its 
signature from $+$ to $-$ for increasing $\Theta_-$. 
Then, again the number of nodes, $N_-$, 
increases by one when $\Theta_-$ crosses $(2 n+1)\pi$ 
in the increasing direction. 
For the $\tau>0$ side, we can also give an analogous discussion. 
Using the fact that $q_{\pm}$ is nodeless when 
the curves $\mp C_{\pm}$ start from $P_{\pm}$ with 
$0<\Theta_{\pm}<\pi$ (See appendix), 
the proof is completed. \hfill$\Box$\vs

Now, we are ready to assign a number $N$ for 
each intersection point. We define $N$ by the relation, 
\begin{equation}
 \pi N<\Theta\,(:=\Theta_+ +\Theta_-)\leq\pi(N+1). 
\label{numN}
\end{equation}
Since both $N_+$ and $N_-$ must be positive, 
we have $\Theta_{\pm}>0$. Thus we obtain \vs\newline
{\bf Proposition 7}:
\begin{equation}
 \Theta>0, \quad \mbox{and} \quad N\ge 0. 
\end{equation}

\subsection{a Proof of the Theorem 1}
\label{subN3}

In this subsection, we give a proof of Theorem 1. 
We assume $\dot\phi$ stays positive 
throughout this subsection. 

First, we explain the fact that 
the number of negative modes at an intersection point $P$ 
is not simply given by 
the summation of 
numbers of nodes of $q_{\pm}$, i.e.,
$N_- +N_+$.  
In order to obtain the number of negative modes, 
the information about $N_+$ and $N_-$ must 
be supplemented with the knowledge of the signature of 
\begin{equation}
 B:={\dot {\bar q}_+ \over \bar q_+}
  -{\dot {\bar q}_- \over \bar q_-}.
\label{defB}
\end{equation}
{\bf Proposition 8}:
{\it The number of negative modes is given by }
\begin{eqnarray*}
\begin{array}{lcc}
N_- +N_+ +1, & \mbox{for} & B>0, \\
N_- +N_+, & \mbox{for} & B\le 0. \\
\end{array}
\end{eqnarray*}
\newline{\it Proof.}
We define  $\bar q_{\pm}(\lambda_j)$ by the solution 
of Eq.~(\ref{yeq}) that satisfies 
regular boundary conditions imposed at 
one boundary $\tau\to \tau_{\pm}$. 
Then, the quantity corresponding to 
$B$ can also be defined for $\lambda_j\ne 0$ by 
replacing $\bar q_{\pm}$ with $\bar q_{\pm}(\lambda_j)$ 
in (\ref{defB}). 
We denote it by $B(\lambda_j)$. 
In the present case, the potential $U$ is non-singular. 
Then, if we gradually decrease $\lambda_j$, $B(\lambda_j)$
also decreases monotonically 
until it diverges to $-\infty$. When $B(\lambda_j)$ diverges, 
either 
$\bar q_+(\lambda_j)$ or $\bar q_-(\lambda_j)$ vanishes, 
and hence the number of nodes changes. 

In the case of $B>0$, decreasing $\lambda_j$ from zero, 
$B(\lambda_j)$ vanishes at some 
$\lambda_j$ before the number of nodes of $q_{\pm}(\lambda_j)$ 
changes. 
This $\lambda_j$ is the largest negative eigen value, and 
the corresponding eigen function still has $N_- +N_+$ nodes. 
Hence, the number of negative modes is $N_- +N_+ +1$. 
On the other hand, when $B\le 0$, we gradually 
increase $\lambda_j$. Then, we have the first eigen 
mode with positive $\lambda_j$ which has $N_- +N_+$ nodes. 
Hence, the number of negative modes is $N_- +N_+$. 
\hfill$\Box$\vs

\vspace{5mm}
\begin{figure}[htb]
\centerline{\epsfxsize5cm\epsfbox{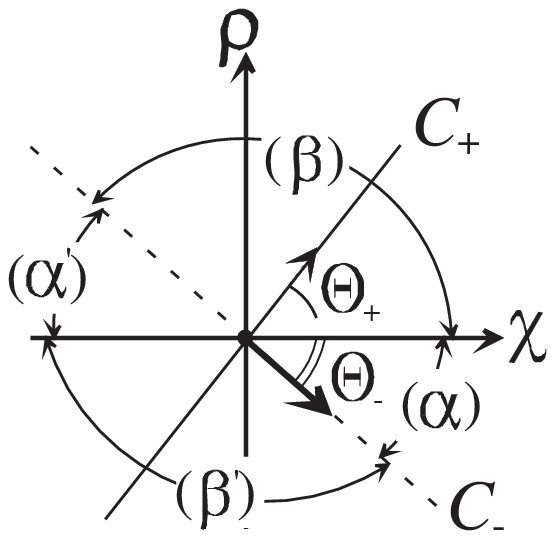}}
\vspace{3mm}
\refstepcounter{figure}
{\small FIG.\thefigure.
The diagram is rotated so that the $\chi$-direction points 
rightward. For a fixed direction of the curve $C_-$, the 
direction of the curve ${C}_+$ is classified into four regions. 
The boundaries of four regions belong to the region with 
the smaller $\Theta_+$. }
\addcontentsline{lof}{figure}
{\protect\numberline{\thefigure}{fig4}}
\label{fig4}
\end{figure}
\vspace{5mm}

{}From Proposition 8, we find that to show Theorem 1 
is equivalent to show 
the following Proposition.\vs\newline
{\bf Theorem 1'}:\newline
{\it $B>0$ $\Leftrightarrow$ $\pi(N_+ + N_- +1) 
<\Theta(:=\Theta_+ +\Theta_-)\leq\pi(N_+ + N_- +2)$ 
$($case $(a))$, and $B<0$ $\Leftrightarrow$ $\pi(N_+ + N_-) 
<\Theta \leq\pi(N_+ + N_- +1)$ $($case $(b))$, 
where case $(a)$ includes the case in which $B$ diverges. }
\vs\newline{\it Proof.}
We first consider which case 
is realized for given $\Theta_{\pm}$. 
For example, we fix $\Theta_-$ as shown in Fig.\ref{fig4}, which is 
a close-up view around an intersection point in the 
$(\bar\phi,\bar\Pi)$-diagram as shown in Fig.\ref{fig2}$(a)$.  
This figure is rotated so as for the $\chi$-direction to point 
rightward.
Then, we classify 
the possible direction of the curve ${C}_+$ into 
four regions labeled 
with $(\alpha)$, $(\beta)$, $(\alpha')$ and $(\beta')$. 
The boundaries of four regions belong to the region with 
the smaller $\Theta_+$. 
Then, we can see 
that case (a), in which $\pi(N_+ + N_- +1) 
<\Theta(:=\Theta_+ +\Theta_-)\leq\pi(N_+ + N_- +2)$, 
is realized when ${\Theta}_+$ is 
in the region $(\alpha)$ or $(\alpha')$. 
To show $B\geq 0$ in this case, 
we evaluate $B$ by using Eq.~(\ref{pim}) to find 
\begin{equation}
 B={1\over \bar q_- \bar q_+}\left(
    \bar\pi^{\varphi}_{+}\bar\varphi_{-}
   -\bar\pi^{\varphi}_{-}\bar\varphi_{+}\right). 
\label{Beq}
\end{equation}
In the regions $(\alpha)$ and $(\beta)$, the quantity in the round 
bracket is positive. 
On the other hand, the quantity $\bar q_- \bar q_+$ is positive in 
the regions $(\alpha)$ and $(\beta')$. 
Consequently, we obtain $B\geq 0$ in the regions 
$(\alpha)$ and $(\alpha')$. 
The case in which the curve $C_-$ is pointing upward can be 
discussed in the same way. \hfill$\Box$\vs\newline
{\bf Remark}:
The value of $\Theta$ at $P$ does not cross $n\pi$ under a continuous 
deformation of the curves $C_{\pm}$ as long as 
the way of intersection at $P$ is unchanged 
and the curves do not cross the points $O$, $P_+$ and $P_-$. 
Hence, the number $N$ at a point $P$ is invariant 
under such a continuous deformation of $C_{\pm}$. 

\section{multiplicity of walls}
\label{oldsec6}
From our method to find 
solutions from the intersection points of two curves $C_{\pm}$, 
the solution may have several nodes of $\dot\phi$. 
We show this number of nodes, $M$, can also be read from 
the $(\bar\phi,\bar\Pi)$-diagram. 

For this purpose, we prove the following statement.\vs\newline 
{\bf Proposition 9}:
{\it A point where $\dot\phi_{\pm}(\phi_{\pm}^i)=0$ 
appears only from the boundary specified by $\tau=0$, 
as we vary $\phi_{\pm}^i$ along the curves $C_{\pm}$.}
\vs\newline
{\it Proof.} 
Let us assume that a point at which 
$\dot\phi_{\pm}(\phi_{\pm}^i;\tau)=0$ 
appeared at an intermediate point $\tau=\tau_s\ne 0$ abruptly. 
Then, we must have $\ddot\phi_{\pm}(\phi_{\pm}^i;\tau_s)=0$ 
from the continuity 
of $\phi_{\pm}(\phi_{\pm}^i;\tau_s)$ with respect 
to both $\phi_{\pm}^i$ and $\tau$. 
However, the field equation (\ref{feq}) 
indicates that $dV/d\phi=0$ at this point. 
Then, this solution must be one of the trivial 
solutions $\phi_{\pm}\equiv\Phi_{\pm}$ or 
$\phi_{\pm}\equiv 0$, which correspond 
not to an intermediate point of $C_{\pm}$ but to their end 
points. This is a contradiction. \hfill$\Box$\vs

\noindent
{\bf Theorem 2}: \newline
{\it The number of nodes of 
$\dot\phi(\tau)$ for a solution corresponding to an 
intersection point $P$ is given by }
\begin{equation}
 M=(\chi_{-} - \chi_{+})/ \pi|_{{\rm at} \,P}. 
\label{numM}
\end{equation}
{\it Proof.}
From Proposition 9, the number of nodes of 
$\dot\phi_{\pm}(\phi_{\pm}^i;\tau)$ changes only 
when $\dot{\bar\phi}_{\pm}(\phi_{\pm}^i)=0$ 
as we change $\phi_{\pm}^i$ along the curves $C_{\pm}$. 
This happens if and only if 
$\chi_{\pm}=m \pi$ where $m$ is an integer. 
Denote by $M_{\pm}$ the number of nodes of $\dot\phi_{\pm}$ 
$($existing between $0$ and $\tau_{\pm}$ 
excluding the point $\tau=\tau_{\pm})$.  
To avoid duplicative counting, we do not count the point 
$\tau=0$ as a node when $\phi_-(0)=0$ but we count it as 
a node when $\phi_+(0)=0$.  
Here we use the same technique that was used to count the 
number of nodes of $x_{\pm}$ in the subsection \ref{subN2}. 
Corresponding to Eq.(\ref{xbardot}), 
the derivative of $\dot\phi_{\pm}$ with respect to $\tau$ at 
$\tau=0$ is calculated as 
\begin{equation}
 \left.{d\over d\tau} \dot\phi_{\pm}\right\vert_{\tau=0}
  = {dV\over d\phi}(\bar\phi_{\pm}),\quad (\mbox{when}~ 
 \dot{\bar \phi}_{\pm}=0).  
\end{equation}
This quantity is positive (negative) when $\bar\phi_{\pm}$ is negative 
(positive). 
Then, for the same reasoning as was used 
to determine the number of nodes for $q_{\pm}(\tau)$, 
we find the relations, 
\begin{eqnarray}
 && \pi (M_{-}-1) < \chi_- \leq \pi M_{-}, \cr
 && \pi M_{+} \leq -\chi_+ < \pi (M_{+}+1).  
\label{Mbound}
\end{eqnarray}
For an intersection point, $\phi(\tau)$ is given by 
$\phi_-(\tau)$ for $\tau<0$ and 
$\phi_+(\tau)$ for $\tau\geq 0$. 
Thus, the number M, which is the total number of 
nodes of $\dot\phi(\tau)$, is given by $M_+ + M_-$. 
Then, from Eq.(\ref{Mbound}), 
we find $\pi(M-1)<$ $\chi_- -\chi_+<$ $\pi(M+1)$. 
Since $\chi_- -\chi_+$ 
must be devided by $2\pi$, we finally find $M$ is given 
by (\ref{numM}). \hfill$\Box$\vs\newline
{\bf Remark}:
Theorem 2 tells that $M/2$ is the winding number 
around $O$ of the continuous curve $P_- P P_+$ 
that is obtained by connecting $P_-, P$ and $P_+$ 
with $C_-$ and $C_+$. 
Thus, the number $M$ is invariant under a continuous deformation of 
curves $C_{\pm}$ as long as the curves do not cross the points $O$, 
$P_+$ and $P_-$. 

\section{Comparison of the values of action 
between various CD bounce solutions}
\label{oldsec3}

We introduce a pseudo-action defined 
as a function of $\phi^i_{-}$ and $\phi^i_{+}$ by  
\begin{equation}
 \tilde S_E(\phi^i_{+},\phi^i_{-})=S_+(\phi^i_{+})+S_-(\phi^i_{-}),
\end{equation}
and 
\endmulticols
%\vspace{5mm}
\begin{figure}[htb]
\centerline{\epsfxsize12cm\epsfbox{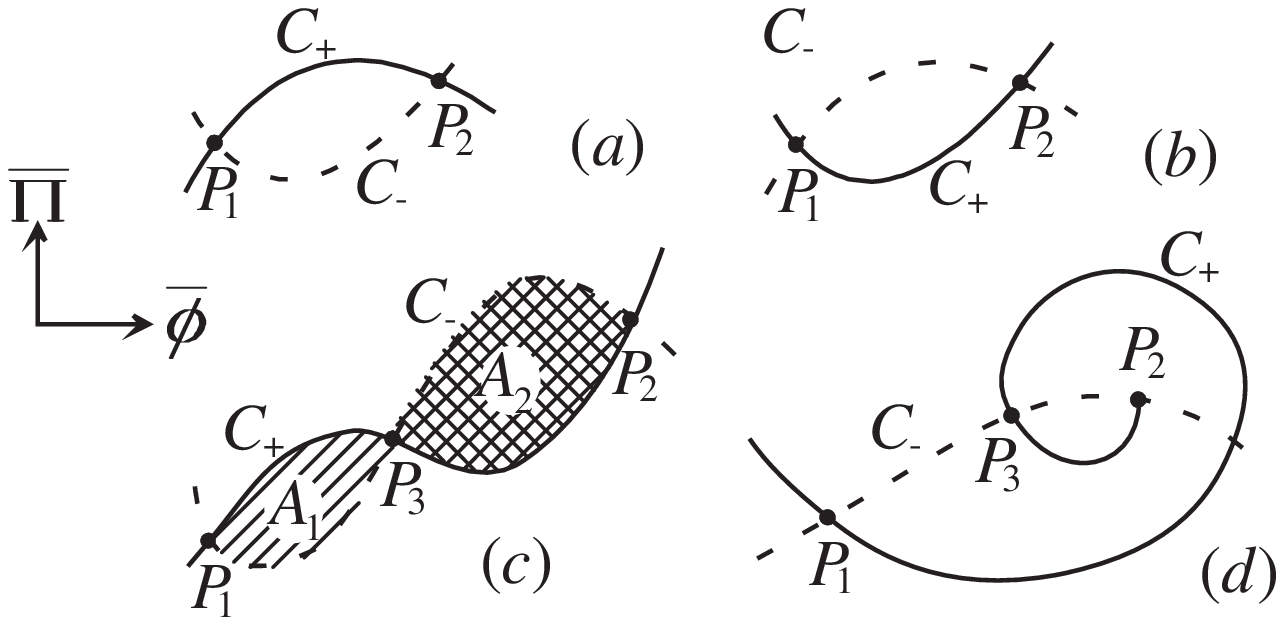}}
\vspace{3mm}
\caption{A few examples to show various ways of connection 
between two intersection points $P_1$ and $P_2$. 
It is possible to compare the values of action from this diagram,  
as explained in the text. We can show that  
$S_E[P_1]>S_E[P_2]$ for case $(a)$, $S_E[P_1]<S_E[P_2]$ for case $(b)$, 
$S_E[P_1]<S_E[P_2]$ if $A1<A2$ for case $(c)$, and  
$S_E[P_1]<S_E[P_2]$ for case $(d)$. }
\label{fig5}
\end{figure}
\vspace{5mm}
\multicols{2}
\begin{equation}
 S_{\pm}(\phi^i_{\pm})=\pm 2\pi^2\int_0^{\tau_{\pm}} 
   d\tau\, L[a_{\pm}(\phi^i;\tau),\phi_{\pm}(\phi^i;\tau)], 
\end{equation}
where 
\begin{equation}
 L[a,\phi]= 
   a^3 \left[{1\over 2}\dot\phi^2+V(\phi) 
     -{3\over \kappa}
    \left(\left({\dot a\over a}\right)^2
      +{1\over a^2}\right)\right]. 
\label{partaction}
\end{equation}
Note that $S_{\pm}(\phi^i_{\pm})$ is 
defined along the curve $C_{\pm}$, and hence 
so $\tilde S_E$ is. To the contrary, the original 
action $S_E$ is well-defined only at intersection points which 
represent CD bounce solutions. 
For CD bounce solutions, i.e., at intersection points, 
we have $\tilde S_E=S_E$.

Then, we calculate 
\begin{eqnarray}
 {d S_{\pm}\over d\phi_{\pm}^i} & = & 
 \pm 2\pi^2 \Biggl\{\int_0^{\tau_{\pm}} d\tau 
\Bigl( {\delta L[a_{\pm},\phi_{\pm}]\over \delta a_{\pm}} 
 {\partial a_{\pm}\over \partial\phi_{\pm}^i}
\cr &&\quad\quad\quad\quad\quad
   + {\delta L[a_{\pm},\phi_{\pm}]\over \delta \phi_{\pm}} 
 {\partial \phi_{\pm}\over \partial\phi_{\pm}^i}\Bigr)
\cr
 &&\quad\quad\quad +\left[
   -{6\over \kappa} a_{\pm}\dot a_{\pm}
           {\partial a_{\pm}\over \partial\phi_{\pm}^i}
   +a^3_{\pm}\dot\phi_{\pm}{\partial \phi_{\pm}\over 
    \partial\phi_{\pm}^i}
  \right]_{0}^{\tau_{\pm}}
\cr &&\quad\quad\quad +
    {d\tau_{\pm}(\phi^i)\over d\phi_{\pm}^i}
    L[a_{\pm},\phi_{\pm}]|_{\tau=\tau_{\pm}} \Biggr\} \cr
& = & 
 \mp 2\pi^2 \bar\Pi_{\pm} \bar\varphi_{\pm}, 
\end{eqnarray}
where partial differentiations with respect to 
$\phi_{\pm}^i$ are taken for fixed $\tau$, 
and we used the fact $(\phi_{\pm},a_{\pm})$ satisfies the background 
equations, i.e., 
\begin{eqnarray}
&& {\delta L\over\delta a} :=  
   -{d\over d\tau}\left({\partial L\over \partial \dot a}\right)+ 
         {\partial L\over \partial a} =0, 
\cr && 
{\delta L\over \delta \phi} := 
    -{d\over d\tau}\left({\partial L\over \partial \dot\phi}\right)+ 
         {\partial L\over \partial \phi} =0. 
\end{eqnarray}
Integrating this expression, we find that the difference 
of the values of action between two bounce solutions 
specified by points $P_1$ and $P_2$ is given by 
\begin{equation}
 S_E[P_2]-S_E[P_1]=2\pi^2 \int_{P_1}^{P_2} 
   d\bar\phi (\bar\Pi_{-} -\bar\Pi_{+}). 
\end{equation}
When the two curves connecting $P_1$ and $P_2$ 
do not intersect with each\, other, the\, signature\, of\, 
$S_E[P_2]-S_E[P_1]$ is totally determined 
by\, the\, topological\, information\, of 
\vspace{5mm}
\begin{figure}[bht]
\centerline{\epsfxsize6cm\epsfbox{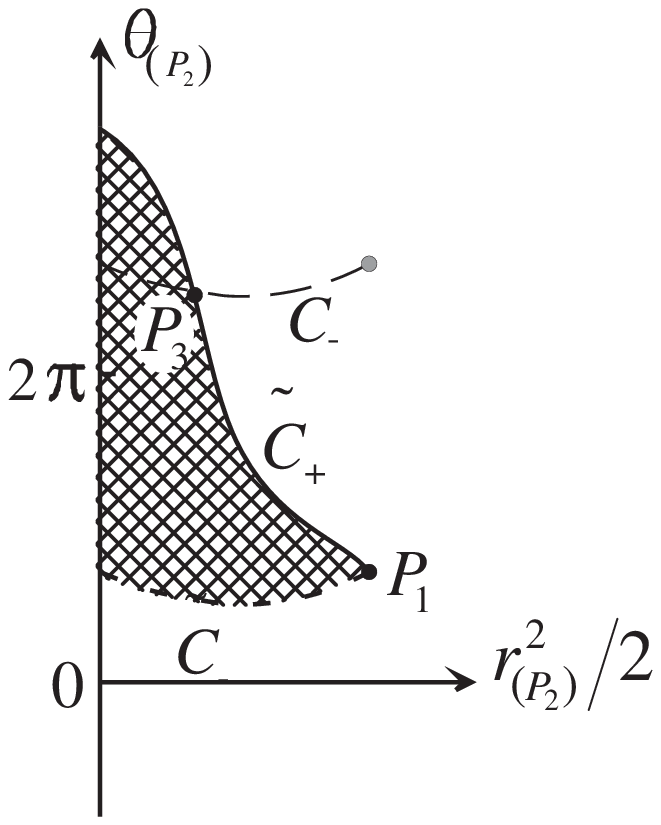}}
\vspace{3mm}
\refstepcounter{figure}
{\small FIG.\thefigure.
A plot of $C_{\pm}$ connecting $P_1$ and $P_2$ 
in the spherical coordinate $(r_{(P_2)},\theta_{(P_2)})$ 
centered at $P_2$. 
In this coordinate 
$C_{\pm}$ has many copies in this coordinate.  
We specify one pair of curves by the condition 
that the point $P_1$ exists in the region $0\leq\theta<2\pi$. 
We attach $\tilde{}$ to these specific curves 
as $\tilde C_{\pm}$ to distinguish them from other copies. 
Then, in this coordinate, the area surrounded by $\tilde C_+$ 
and $\tilde C_-$ gives the difference of the values of 
action at $P_1$ and $P_2$.}
\addcontentsline{lof}{figure}
{\protect\numberline{\thefigure}{fig6}}
\label{fig6}
\end{figure}
\vspace{5mm}
\noindent
the connecting curves. 
As shown in Fig.\ref{fig5}$(a)$, if the region surrounded by the two curves 
is on the left hand side when we move from $P_1$ 
to $P_2$ along the curve $C_-$,  
$S_E[P_1]>S_E[P_2]$ can be concluded. 
The difference is proportional to the 
area surrounded by the two curves. 
In the opposite case as shown in Fig.\ref{fig5}$(b)$, 
we can say $S_E[P_1]<S_E[P_2]$. 
In such a case as presented in Fig.\ref{fig5}$(c)$, we need to compare 
the areas $A_1$ and $A_2$ to determine which point has 
the smaller value of action. 
In some cases, however, 
even when there is another intersection point 
between $P_1$ and $P_2$ 
as shown in Fig.\ref{fig5}$(d)$, we can determine 
which point has the smaller value of action
from the topological information alone. 
In the case of Fig.\ref{fig5}$(d)$, we can say that $S_E[P_1]<S_E[P_3]$ 
and $S_E[P_3]<S_E[P_2]$. 
Hence $S_E[P_1]<S_E[P_2]$ is concluded.

{}For our present purpose, we do not have to give a general statement 
in what situation we can determine the signature of 
$S_E[P_2]-S_E[P_1]$ from the topological information alone. 
Only the case discussed below is of special importance. 
Let us draw the two curves $C_{\pm}$ by using 
a spherical coordinate $(r_{(P_2)},\theta_{(P_2)})$ whose 
origin is located on $P_2$, i.e., the point $P_2$ 
corresponds to the line $r_{(P_2)}=0$. 
In this coordinate, when the value of $\theta_{(P)}$ of two points 
are different by $2n\pi$ with an integer $n$, 
these two points are identical on the original 
$(\bar\phi,\bar\Pi)$-plane. 
Thus, there are many copies of the curves $C_{\pm}$ 
in the $(r_{(P_2)},\theta_{(P_2)})$-coordinate. 
We choose one pair of $C_{\pm}$ both of which pass through 
a common point corresponding to $P_1$ 
in the $(r_{(P_2)},\theta_{(P_2)})$-coordinate, and 
denote them by $\tilde C_{\pm}$. 
If we choose this point, say, to satisfy 
$P_1$ with $0<\theta_{(P_2)}<2\pi$, 
the two curves $\tilde C_\pm$ 
can be drawn without any ambiguity. 
If the two curves $\tilde C_{\pm}$ 
do not intersect between $P_1$ and $P_2$ in this coordinate as shown in 
Fig.\ref{fig6}, where we used $r_2^2/2$ as the horizontal coordinate 
instead of $r_2$, the $(S_E[P_2]-S_E[P_1])/2\pi^2$ is given by the area 
of the shaded region, and we can say it is positive. 
We note that no intersection in this coordinate 
does not mean no intersection on the original 
$(\bar\phi,\bar\Pi)$-plane. 
In fact, Fig.\ref{fig6} represents the same situation that 
was shown in Fig.\ref{fig5}$(d)$. 

The result obtained in this section is summarized by 
the following Theorem. 
\vs\newline
{\bf Theorem 3}:\newline
{\it 
We choose one pair of $C_{\pm}$ 
such that they pass through 
a common point corresponding to $P_1$ 
in the $(r_{(P_2)},\theta_{(P_2)})$-coordinate, 
and denote them by $\tilde C_{\pm}$ 
as shown in Fig.\ref{fig6}. 
If these two curves 
does not intersect between $P_1$ and $P_2$ and 
if the value of $\theta_{(P_2)}$ at $r_{(2)}=0$ 
on $\tilde C_+$ is larger than 
that on $\tilde C_-$, then $S_E[P_1]<S_E[P_2]$.}\vs\newline
{\bf Remark}:
Theorem 3 depends only on the topological 
information about intersections between the curves $C_+$ 
and $C_-$. Hence, we can apply the same statement for 
any deformed diagram which preserves this topological 
information. 

\section{a proof of 
the ``no-negative mode theorem''}
\label{proof}

Now, we prove the Main Theorem ( 
the ``no-negative mode theorem''). 
Namely, we prove if an intersection point $P$ has 
either negative modes or nodes of $\dot\phi(\tau)$, 
there exists another intersection point $P'$ 
which has a smaller value of action than $P$. 
By using Theorem 1 given in 
Sec.\ref{oldsec5} and Theorem 2 
given in Sec.\ref{oldsec6}, 
the Main Theorem is reduced to the following statement. 
\vs\newline
{\bf The Main Theorem'}:\newline
{\it The intersection point with $N\ne 0$ or $M\ne 0$
cannot be the solution that realizes the minimum 
value of action.}

\endmulticols

%\vspace{5mm}
\begin{figure}[htb]
\centerline{\epsfxsize13cm\epsfbox{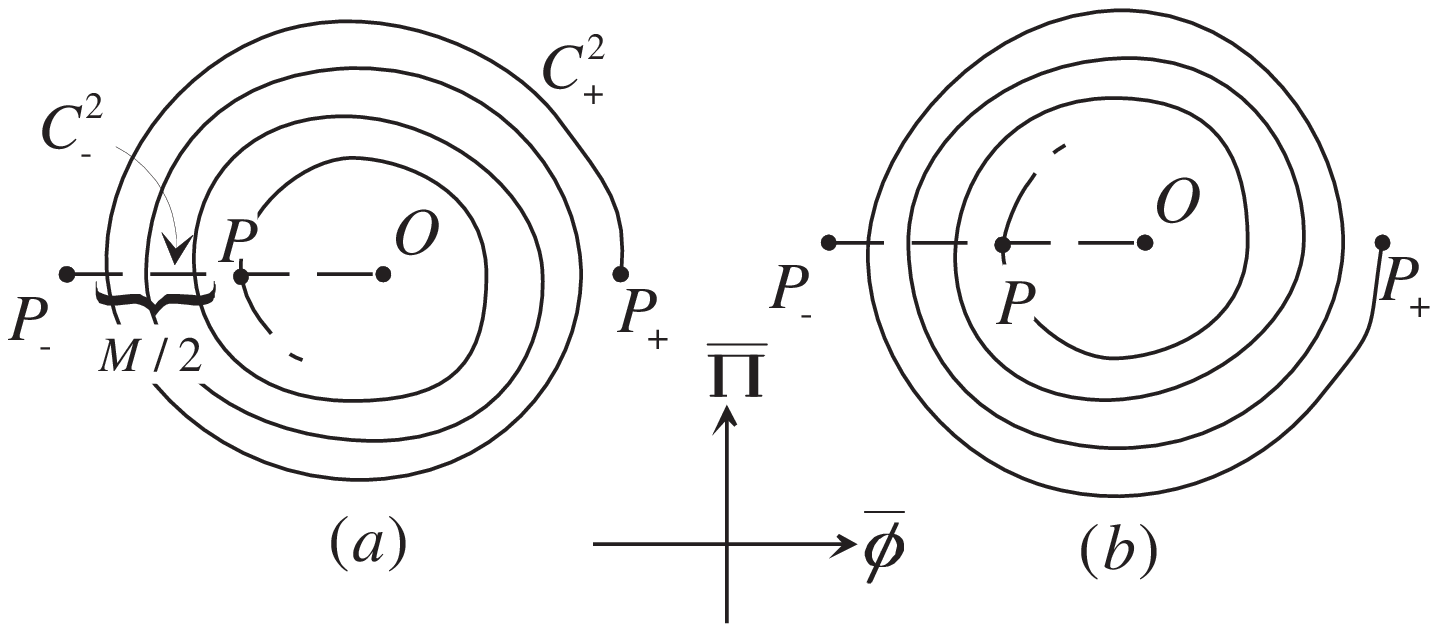}}
\vspace{3mm}
\caption{Under the constraint that the curves do not cross the 
points $O$, $P_+$ and $P_-$, 
the deformed diagram $D^2$ consisting of $C_{\pm}^2$ 
can be reduced to one of these two diagrams. 
However, we can see that $M<0$ for the point $P$ in 
the diagram $(b)$. As we know $M\ge 0$, 
the possibility $(b)$ is removed. 
In the diagram $(a)$, the winding numbers $M/2$ can be zero.}
\label{fig7}
\end{figure}
%\vspace{5mm}

\multicols{2}
To show the above statement, we first define curves $C^1_{\pm}$ 
by deforming $C_{\pm}$ continuously to make $C_{-}$ to be a 
straight line connecting $P_-$ and $O$. 
Under this deformation of curves, the way of intersections, 
i.e., the value of $\Theta:=\Theta_+ + \Theta_-$ 
at each intersection point, 
is kept unaltered, and the curves do not cross the points $O$, 
$P_+$ and $P_-$. 
The diagram composed of these deformed curves 
keeps the numbers $N$ and $M$ assigned for each intersection point, 
and also it still keeps some information about the difference 
between the values of action for different points. 
In fact, under the situation such that 
Theorem 3 determines which point 
has the smaller value of action, we can apply the Theorem 
to this deformed diagram in place of the original one. 
We denote this diagram consisting of $C^1_{\pm}$ by $D^1$. 

Focusing on a point $P$ in $D^1$, we further 
deform $C_+^1$ continuously. Under this deformation, 
we allow the way of connections 
at points other than $P$ to change, 
and hence the number of intersection points can be varies. 
Furthermore, we allow the value of $\Theta$ at $P$ to deviate 
from $\Theta_{(P)}$, where $\Theta_{(P)}$ is the original value of 
$\Theta$ at $P$. 
However, we do not allow the curve to 
cross the points $O$, $P_+$ and $P_-$.
Hence, the number $M$ for $P$ 
is kept unaltered.  
Then, the diagram can be reduced to 
the diagram that are shown in Fig.\ref{fig7}$(a)$. 
The cases given by Fig.\ref{fig7}$(b)$ are not realized 
because the number $M$ for the point $P$ 
becomes negative. 
We denote these further deformed 
curves by $C^2_{\pm}$ (c.f., $C^2_- =C^1_-)$ 
and the corresponding diagram by $D^2$. 
As noted in the previous section, the number $M$ 
represents twice the number of windings of the curve 
$P_- P P_+$ around $O$. 

\begin{figure}[htb]
\centerline{\epsfxsize13cm\epsfbox{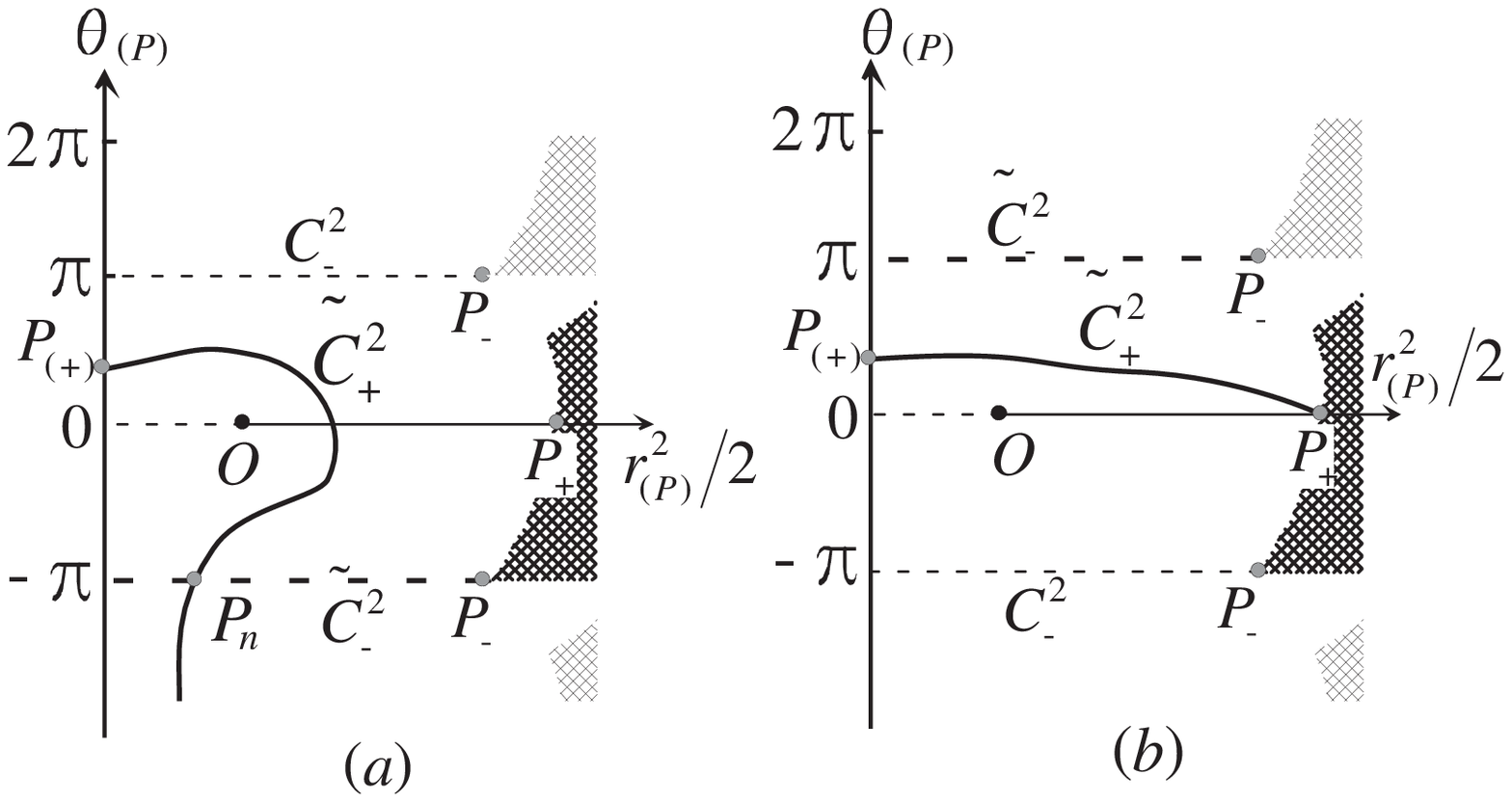}}
\vspace{3mm}
\caption{$(a)$ The distorted diagram $D^2$ in the 
$(r_{(P)},\theta_{(P)})$-coordinate corresponding to 
Fig.7$(a)$ when $M\ne 0$. The curve $C_+^2$ is shown in 
from $P$ in the direction for $P_+$. 
The shaded region corresponds to the 
forbidden regions labeled with $(\alpha)$ and $(\gamma)$ 
in Fig.2$(b)$. 
$P_n$ is the intersection point neighboring to $P$ in $D^2$. 
$(b)$ The same diagram when $M=0$. }
\label{fig8}
\end{figure}

Now we use the $(r_{(P)},\theta_{(P)})$-coordinate 
introduced in Sec.\ref{oldsec3}. In this coordinate, 
the curves have many copies. As before, 
we attach $~\tilde{}~$ to a specific pair of curves to 
distinguish it from others. It is explained below how 
we select the specific curves. 
The deformed diagram $D^2$ in this coordinate is given by 
Fig.\ref{fig8}$(a)$ when $M\ne 0$ and by 
Fig.\ref{fig8}$(b)$ when $M=0$. 
In both cases, $\tilde C^2_+$ and $\tilde C^2_-$ are 
shown by the thick rigid line and by 
the thick dashed line, respectively. 
Here, $P_n$ in Fig.\ref{fig8}$(a)$ is 
the intersection point neighboring to $P$ 
in Fig.\ref{fig7} $(a)$. 
Later, we consider the process that the diagram 
$D^1$ is recovered from $D^2$ through continuous deformation 
of curves. In the midst of this process, 
we use the notation $\tilde C_{\pm}$ 
to specify the curves that are denoted by 
$\tilde C_{\pm}^2$ in the diagram $D^2$.  
We refer to the point corresponding 
to the point $P$ on the curve $\tilde C_+$ by $P_{(+)}$, 
(Do not be confused with $P_+$!)  and 
we denote 
the value of $\theta_{(P)}$ at $P_{(+)}$ by $\theta_+$. 
In both case $(a)$ and case $(b)$, we choose 
$\tilde C_+^2$ so as to satisfy $0<\theta_{+}<\pi$. 
The part of the curve $C_-^2$ 
connecting $P_-$ and $P$ is given by  
the lines with $\theta_{(P)}=\pi(2n-1)$, 
where $n$ is an integer. 
We choose the line with $n=0$ as $\tilde C_-^2$ for case $(a)$, 
and that with $n=1$ for case $(b)$. 
They are presented by the thick dashed lines in Fig.\ref{fig8}. 
\vs\newline
{\bf Proposition 10}:
{\it We consider the diagram $D^1$ that is 
recovered from the diagram $D^2$ given in Fig.\ref{fig8}$(a)$ or $(b)$. 
For this diagram, $[\theta_+/\pi]$ 
is given by $[\Theta_{(P)}/\pi]$, where 
$\Theta_{(P)}$ is the value of $\Theta$ at $P$ in the original diagram.
$($Here we denoted the largest integer less than or equal to $x$ 
by $[x].)$ 
Furthermore, $\theta_{+}$ in the diagram $D^1$ 
must be greater than $0$. }
%\vs\newline

\endmulticols
%\vspace{5mm}

\multicols{2}
\noindent
{\it Proof.}
From Fig.\ref{fig7}$(a)$, one can see that the value of $\Theta$ at $P$ 
for this deformed diagram $D^2$ is also in between $0$ and $\pi$. 
Hence, we find $[\theta_{+}/\pi]=[\Theta/\pi]$ for the 
diagram $D^2$.  
In recovering the diagram $D^1$ from $D^2$, the 
direction of the curve $C_+^2$ 
in Fig.\ref{fig7}$(a)$ may rotate around $P$.  
Under this rewinding, both $\Theta$ and $\theta_+$ 
increase when the direction of $C_+$ 
rotates in the the anti-clockwise direction, and 
both of them are divided by $\pi$ when 
$C_+$ becomes parallel to $C_-$ at $P$. 
Hence, we can say that the relation 
$[\Theta/\pi]=[\theta_+/\pi]$ is maintained 
under the deformation from $D^2$ to $D^1$. 
Therefore, when the diagram $D^1$ is recovered, 
the point $P_{(+)}$ is moved along the $\theta_{(P)}$-axis 
to a point satisfying $[\theta_{(P)}/\pi]=[\Theta_{(P)}/\pi]$. 
The latter part of the Proposition is manifest from 
Proposition 7, which tells 
$\Theta_{(P)}$ must be positive. \hfill$\Box$\vs

\noindent
{\bf Theorem 4}:\newline
{\it A point $P$ with $M\ne 0$ 
cannot be the solution that realizes the minimum 
value of action. }
\vs\newline
{\it Proof.} 
Let us consider the continuous deformation of curves 
to recover the diagram $D^1$ from $D^2$ 
in the $(r_{(P)},\theta_{(P)})$-coordinate. 
As shown in Fig.\ref{fig8}$(a)$, 
$\tilde C^2_+$ has an intersection point $P_n$
with $\tilde C_-^2 (=\tilde C_-^1)$. 
From the continuity of deformation, 
the curve $\tilde C^1_{+}$, as well as $\tilde C^2_+$, must intersect 
with $\tilde C^1_-$. 
Then, let us refer to 
the first intersection point nearest to $P$ as $P'$. 
Since the curve $\tilde C^1_+$ does not 
intersect with the curve $\tilde C^1_-$ between $P$ and $P'$ by definition, 
the area surrounded by these 
two curves connecting $P$ and $P'$ has a definite signature.  
By virtue of Proposition 10, 
$P_{(+)}$ in the diagram $D^1$ must be on the upper side of 
the $\theta_{(P)}=0$ line. Hence, by using Theorem 3, 
we conclude $S_E[P']< S_E[P]$. \hfill$\Box$\vs

\noindent
{\bf Theorem 5}:\newline
{\it A point $P$ with $N\ne 0$ and $M=0$
cannot be the solution that realizes the minimum value of action. }
\vs\newline
{\it Proof.} 
Again, we consider the continuous deformation 
to recover the diagram $D^1$ from $D^2$ 
in the $(r_{(P)},\theta_{(P)})$-coordinate. 
In the case $M=0$, the diagram $D^2$ is given 
by Fig.\ref{fig8}$(b)$. 
If $N\geq 1$, $\Theta_{(P)}$ must be greater than $\pi$. 
Then, owing to Proposition 10, $\theta_+$ for 
the diagram $D^1$ must be greater than $\pi$.  
Consequently, we find that 
the point $P_{(+)}$ moves to the 
upper side of the $\theta_{(P)}=\pi$ line, while 
the point $P_{+}$ must stay on the opposite side.   
Recalling the forbidden region shown as shaded 
regions in Fig.\ref{fig8}$(b)$, 
we find that the curve $\tilde C_+^1$ must have 
at least one intersection 
point with $\tilde C_-^1$. 
Let us refer to 
the first intersection point nearest to $P$ as $P'$. 
Then, as before, we can use Theorem 3 to 
conclude $S_E[P']< S_E[P]$. \hfill$\Box$\vs

Now, from Theorem 4 and Theorem 5, 
the proof of the Main Theorem is completed. 

\section{summary}

We gave a proof of the ``no-negative mode theorem'', 
which tells that the bounce solution realizing 
the smallest value of action among all $O(4)$-symmetric CD bounce solutions 
has no-negative mode and has no node in $\dot\phi$. 
We summarize the outline of the proof presented in this paper. 
In Sec.III, we introduced a diagram which consists of two curves. 
We showed that each intersection point of these two curves 
corresponds to an $O(4)$-symmetric CD bounce solution. 
In Sec.IV and V, we assigned two numbers $N$ and $M$ for 
each intersection point. The number $N$ is defined by (\ref{numN}), 
and it was shown to represent the number of negative modes when 
$M=0$. 
The number $M$ is defined by (\ref{numM}), and it was shown to be 
the number of nodes of $\dot\phi$. 
As is easily seen from their definitions, 
both $N$ and $M$ have a clear topological meaning. 
In Sec VI, we gave a general rule to compare the values of action 
between different intersection points from the topological 
information of the diagram introduced in Sec.\ref{oldsec2}. 
Collecting these statements related to the topology 
of the diagram, the ``no-negative mode 
theorem'' was proved in sec.VII. 

The ``no-negative mode theorem'' is known to be 
equivalent to the ``no-supercritical 
super curvature mode conjecture''\cite{TSnew}. 
Therefore, we also proved that 
there appears no-supercritical supercurvature 
mode in the one-bubble open inflation universe. 
%The existence of supercritical supercurvature mode 
%results in an exponentially increasing correlation 
%of two-point function of scalar perturbations for 
%large spatial separation. 
%Hence, if it existed, the spectrum of the CMB temperature 
%fluctuations would have large deviation from the standard one. 
%The absence of the supercritical supercurvature mode 
%for general models 
%means that there is a strict constraint on 
%the possible variation of the predicted CMB spectra 
%between various models in the context of one-bubble open inflation 
%scenario. 
%It remains as a future work 
%to study the constraint on the predicted CMB spectrum 
%under the restriction that 
%the supercritical supercurvature mode is absent. 

\vspace{1cm}
\centerline{\bf Acknowledgments}
The author thanks M. Sasaki and J. Garriga for helpful discussions. 
This work was supported in part by Saneyoshi Scholarship.
\vspace{1cm}

\appendix
\section{}
In this appendix, we show that $\mp q_{\pm}(\phi_{\pm}^i;\tau)$ 
is positive in the $\phi_{\pm}^i\to \Phi_{\pm}$ limit. 
For $\phi_{\pm}^i\approx\Phi_{\pm}$, $\phi_{\pm}(\tau)$ will stay near 
$\Phi_{\pm}$, and the geometry does not significantly differ from 
the de Sitter space. In both $+$ and $-$ cases, the limiting 
behavior of  $\mp q_{\pm}(\phi_{\pm}^i;\tau)$ is essentially same. 
To avoid an unnecessary complication, we discuss the false vacuum side 
as a representative case. Similar arguments can be repeated 
for the true vacuum side.   

The equation for 
$\Delta(\phi_-^i;\tau):=\phi_-(\tau)-\Phi_{-}$ can be 
approximately written as 
\begin{equation}
 \ddot\Delta+3H_-\dot\Delta-m^2\Delta=0. 
\label{Deltaeq}
\end{equation}
where $\displaystyle m^2:=\left.
{d^2 V(\Phi)/ d\Phi^2}\right\vert_{\Phi=\Phi_{-}}$. 
Since $H_-$ is almost independent of 
$\Delta$, Eq.~(\ref{Deltaeq}) reduces to a linear 
differential equation. 
Hence, in this limiting case, 
solutions of Eq.~(\ref{Deltaeq}) 
with different initial 
values of $\phi$ are obtained by a simple scaling. 
By introducing a scaling parameter $\epsilon(\phi_-^i)$, 
$\Delta$ 
is written as 
\begin{equation}
 \Delta(\phi_-^i;\tau)=\epsilon(\phi_-^i)\Delta_0(\tau), 
\end{equation}
where $\Delta_0(\tau)$ satisfies Eq.(\ref{Deltaeq}). 
Then, we have 
\begin{equation}
\varphi(\tau)={d\epsilon\over d\phi^i} \Delta_0(\tau),  
\end{equation}
where ${d\epsilon/ d\phi^i}$ is positive, and 
hence $\Delta_0(\tau_-)>0$ because $\varphi(\tau_-)=1$. 
Substituting these approximate expressions into (\ref{qeq}), 
we find
\begin{equation}
 q_-={a_-^2\over 3}{d\epsilon\over d\phi_-^i}
   {\Delta_0^2\over \dot\Delta_0}
   \left(m^2-3H_- \omega -\omega^2\right), 
\label{qmd}
\end{equation}
where $\omega:=\dot\Delta_0/\Delta_0$. 
Since $\Delta_0$ and $\dot\Delta_0$ 
stay non-negative in the present limiting case, 
we have $\omega\geq 0$. 
Furthermore, the equation that $\omega$ should satisfy is given by 
\begin{equation}
 \dot\omega=m^2-3H_-\omega-\omega^2. 
\label{omegadot}
\end{equation}
At $\tau\to\tau_{\pm}$, $\omega\to 0$.  
Then if $\omega$ became greater than $m$, there would be a point 
at which both $\omega=m$ and $\dot\omega\geq 0$ are 
satisfied. However this is impossible from Eq.~(\ref{omegadot}). 
Thus, $\omega<m$ is maintained. 
Then, the condition $0\leq\omega<m$ with Eq.(\ref{qmd}) implies  
that $q_-$ stays positive until $\tau=0$.

\endmulticols

\vspace{4cm}
\begin{thebibliography}{99}
\bibitem{open}
J.R. Gott III, Nature {\bf 295}, 304 (1982);
 J.R. Gott III and T.S. Statler, Phys. Lett. {\bf 136B}, 157 (1984);
 M. Sasaki, T. Tanaka, K. Yamamoto and J. Yokoyama, Phys.
  Lett. B {\bf 317}, 510 (1993); M. Bucher, A.S. Goldhaber and N. Turok, 
  Phys. Rev. {\bf D52} 3314 (1995); 
% M. Bucher, A.  Goldhaber and N. Turok, Nucl. Phys. {\bf B}, 
%   Proc. Suppl. {\bf 43}, 173 (1995); 
   M.  Bucher and N. Turok, Phys.  Rev. {\bf D52}, 5538
  (1995); A.  Linde. Phys. Lett. B351,99 (1995); 
  A. Linde and A. Mezhlumian,
  Phys. Rev. {\bf D52}, 6789 (1995); 
  J. Garcia-Bellido, J. Garriga and X. Montes, Phys. Rev. 
{\bf D57}, 4669 (1998). 
%; A.M. Green and A.R. Liddle,
% astro-ph/9607166.
\bibitem{scmode}
 M. Sasaki, T. Tanaka and K.Yamamoto, Phys. Rev. 
{\bf D51}, 2979 (1995);
% D. Lyth and A. Woszczyna, Phys. Rev. {\bf D52}, 3338 (1995). 
\bibitem{juan}
  J. Garcia-Bellido, A.R. Lidle, D.H. Lyth and D. Wands, Phys. Rev.
{\bf D55}, 4596 (1997).
\bibitem{TSnew}
T. Tanaka and M. Sasaki, Phys. Rev. {\bf D} (1998) {\it in press}. 
\bibitem{TS92}
T. Tanaka and M. Sasaki, Prog. Theor. Phys. {\bf 88} (1992), 503.
\bibitem{Colema}
S. Coleman, Phys. Rev. {\bf D15}, (1977), 2929; 
in {\it The Whys of Subnuclear Physics}, 
Proceedings of the International School, Erice, Italy, 
ed. A. Zichichi, Subnuclear Series Vol.15 
(Plenum, New York. 1979), p.805.
\bibitem{CalCol}
C. G. Callan, Jr. and S. Coleman, Phys. Rev .{\bf D16}, 1762 (1977).
\bibitem{Coluniq}
S. Coleman, Nucl. Phys. {\bf B298}, 178 (1988);
S. Coleman, V. Glaser and A. Martin, Comm. Math. Phys. 
{\bf 58} (1978), 211
\bibitem{ColDeL}
S. Coleman and F. De Luccia, Phys. Rev. {\bf D21}, 3305 (1980). 
\bibitem{dirac}
P.A.M. Dirac, Lectures on Quantum Mechanics, 
Yeshiva University, (1964). 
\bibitem{GXMT1}
J. Garriga, X. Montes, M. Sasaki and T. Tanaka, Nucl. Phys. 
B{\bf 513}, 343 (1998).
\bibitem{HawTur}
S.W. Hawking and N. Turok, Phys. Lett. {\bf B425}, 25 (1998). 
\bibitem{Alex}
A. Vilenkin, Phys. Rev. {\bf D57}, 7069 (1998).
\bibitem{HawMos}
S.W. Hawking and I.G. Moss, Phys. Lett. {\bf 110B}, 35 (1982).
%\bibitem{GXMT2}
%J. Garriga, X. Montes, M. Sasaki and T. Tanaka, 
%{\it in preparation}.
\bibitem{GiHaPe}
G. W. Gibbons, S.W. Hawking and M.J. Perry, Nucl. Phys. 
{\bf B138} (1978), 141.

\end{thebibliography}
\end{document}